\documentclass[graybox, envcountchap]{svmult}

\usepackage{mathptmx}        
\usepackage{amsmath}
\usepackage{amssymb}
\usepackage{color}
\usepackage{helvet}          
\usepackage{courier}         
\usepackage{dirtree}

\usepackage{makeidx}        
\usepackage{graphicx}        
\usepackage{subfig}

\usepackage{multicol}        
\usepackage[bottom]{footmisc}

\usepackage{hyperref}        
\hypersetup{colorlinks=true,urlcolor=blue}

\usepackage[misc]{ifsym}

\newcommand{\be}{\begin{eqnarray}}
\newcommand{\ee}{\end{eqnarray}}

\makeindex             

\begin{document}


\title{Testing Gravity with Black Hole X-Ray Data}

\author{Cosimo Bambi}

\institute{Cosimo Bambi (\Letter) \at Center for Field Theory and Particle Physics and Department of Physics\\Fudan University, 200438 Shanghai, China\\ \email{bambi@fudan.edu.cn}}

\maketitle

\abstract{The analysis of the properties of the X-ray radiation emitted from geometrically thin accretion disks around black holes can be a powerful tool to test General Relativity in the strong field regime. This chapter reviews the state-of-the-art of gravity tests with black hole X-ray data. So far, most efforts have been devoted to test the Kerr hypothesis -- namely that the spacetime around astrophysical black holes is described by the Kerr solution -- and X-ray data can currently provide among the most stringent constraints on possible deviations from the Kerr geometry. As of now, all X-ray analyses are consistent with the predictions of General Relativity.}


\section{Introduction}\label{sec:intro}

In 4-dimensional General Relativity, black holes are completely specified by three parameters: the black hole mass, the black hole spin angular momentum, and the black hole electric charge. This is the result of the celebrated {\it no-hair theorem}, which is actually a family of theorems with specific assumptions and a number of extensions~\cite{Carter:1971zc,Robinson:1975bv,Chrusciel:2012jk}. Uncharged black holes are described by the Kerr solution~\cite{Kerr:1963ud} and have only two parameters: the black hole mass and the black hole spin angular momentum.

The spacetime around astrophysical black holes formed by the complete gravitational collapse of a progenitor astrophysical body should be described well by the Kerr solution. For example, initial deviations from the Kerr metric are quickly radiated away by the emission of gravitational waves at the time of the formation of the black hole~\cite{Price:1971fb}. The presence of nearby stars or of an accretion disk have normally a negligible impact on the spacetime geometry near the black hole event horizon~\cite{Bambi:2017khi,Bambi:2014koa}. The black hole equilibrium electric charge resulting from the difference between the proton and electron masses is tiny and can be ignored for macroscopic objects~\cite{Bambi:2017khi,Bambi:2008hp}. The impact of these and other effects on the spacetime metric around a black hole can be quantified, but it turns out that the induced deviations from the Kerr solution are normally extremely small and negligible even in the case of future very accurate tests of the Kerr hypothesis. On the contrary, macroscopic deviations from the Kerr solution are predicted by a number of scenarios with new physics, from models with macroscopic quantum gravity effects (see, e.g., Refs.~\cite{Dvali:2011aa,Giddings:2014ova}) to scenarios with exotic matter fields (see, e.g., Refs.~\cite{Herdeiro:2014goa,Herdeiro:2016tmi}) or in the case General Relativity is not the correct theory of gravity (see, e.g., Refs.~\cite{Yunes:2009hc,Kleihaus:2011tg}).

\begin{figure}[t]
\centering
\includegraphics[width=0.95\textwidth]{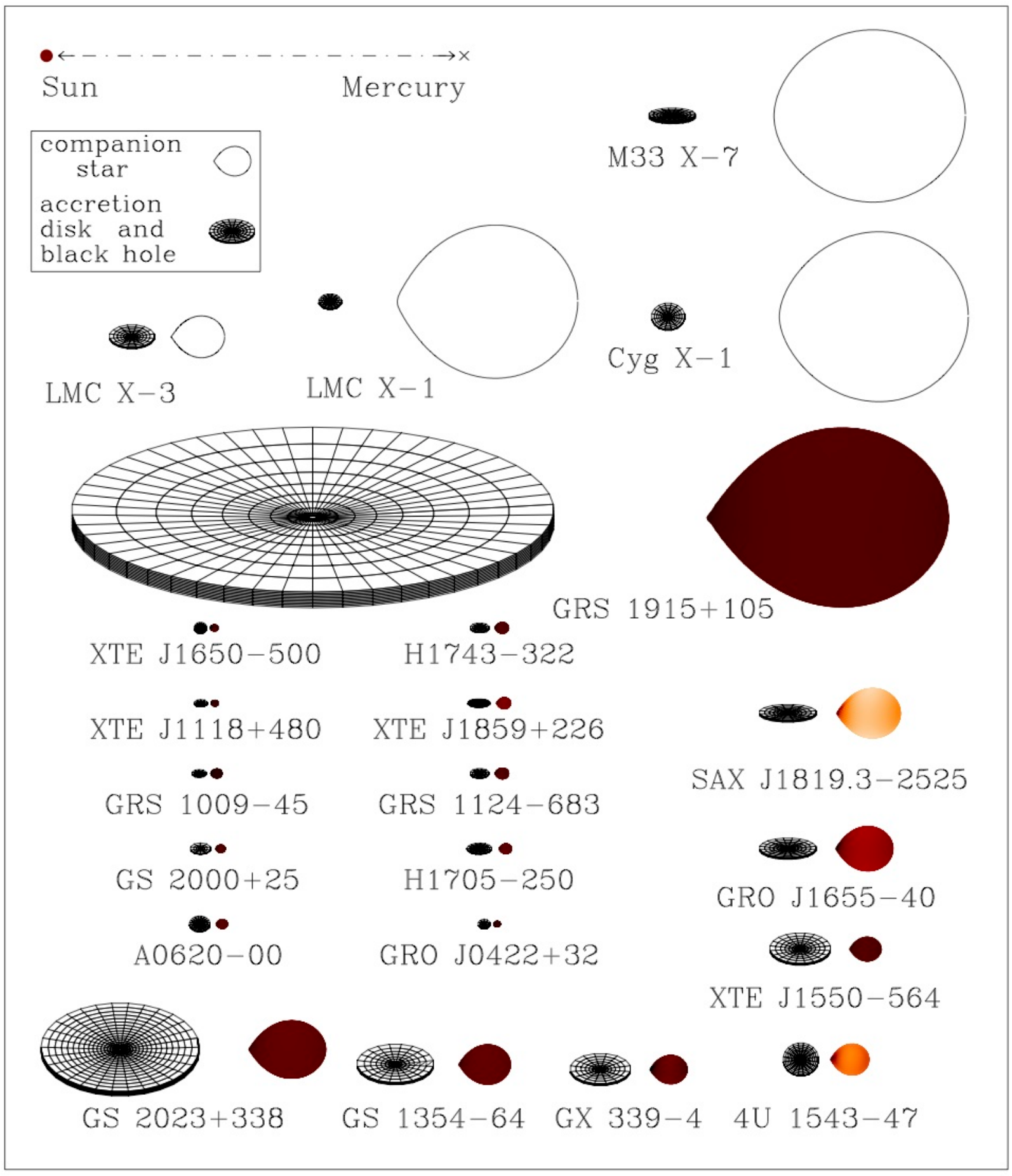}
\caption{Sketch of 22~X-ray binaries with a stellar-mass black hole confirmed by dynamical measurements. For every system, the black hole accretion disk is on the left and the companion star is on the right. The color of the companion star roughly indicates its surface temperature (from brown to white as the temperature increases). We can compare the size of these X-ray binaries with the system Sun-Mercury in the top left corner: the distance Sun-Mercury is about 50~million km and the radius of the Sun is about 0.7~million km. Figure courtesy of Jerome Orosz.}
\label{f-bhxrb}
\end{figure}

From astronomical observations, we know three classes of astrophysical black holes: stellar-mass black holes, supermassive black holes, and intermediate-mass black holes~\cite{Bambi:2017iyh}. 
\begin{itemize}
\item {\it Stellar-mass black holes} are the natural product of the evolution of very heavy stars. When a star exhausts all its nuclear fuel, the thermal pressure of the plasma cannot compensate the gravitational force any longer and the body shrinks to find a new equilibrium configuration. If the quantum pressure of electrons or neutrons can compensate the weight of the collapsing part of the star, we have the formation of, respectively, a white dwarf or a neutron star. If the collapsing body is too heavy, there is no mechanism to stop the collapse and we have the formation of a black hole.

The minimum mass of these black holes is thus set by the Oppenheimer-Volkof limit, which is the maximum mass for a neutron star and is around 2-3~$M_\odot$, depending on the exact matter equation of state, composition, rotation, etc.~\cite{Lattimer:2012nd}. The maximum mass for stellar-mass black holes is probably around 100~$M_\odot$ for objects formed by the direct collapse of primordial metal-poor stars of about 100~$M_\odot$~\cite{Mapelli:2021taw}. For heavier stars, the gravitational collapse is so violent that may destroy the whole system, without leaving any remnant. For stars with higher metallicity, the outer envelope of the star is ejected into space (heavier elements have larger photon cross-sections) and the mass of the final black hole cannot be higher than 20-30~$M_\odot$~\cite{Mapelli:2021taw}.

From stellar evolution studies, we expect a population of 10$^8$-10$^9$~stellar-mass black holes in a galaxy like the Milky Way~\cite{Timmes:1995kp}. While this is a huge number, it is extremely difficult to identify these objects and, as a result, the number of known stellar-mass black holes is much lower. From electromagnetic observations, we currently know about 70~stellar-mass black holes in X-ray binary systems, and only for about 25~objects we have a dynamical measurement of the mass (i.e., from the study of the orbital motion of the companion star we can infer that the mass of the black hole exceeds the Oppenheimer-Volkof limit and therefore it cannot be a neutron star); see Fig.~\ref{f-bhxrb}. The majority of these $\sim$70~stellar-mass black holes are in the Milky Way, while only a few of them are in nearby galaxies.

Most of the known black hole binaries are transient X-ray sources: they are normally in a quiescent state with a very low X-ray luminosity (they can be even too faint to be detected by our X-ray observatories) and sometimes they have an {\it outburst}, when there is a significant transfer of material from the companion star to the black hole. Every year, we may discover 1-3~new black holes, when their binary system has an outburst (see Fig.~\ref{f-bht}). On the other hand, persistent sources are relatively rare: in Fig.~\ref{f-bhxrb}, only Cygnus~X-1, LMC~X-1, LMC~X-3, and M33~X-7 are persistent X-ray sources and all other X-ray binaries are transients. GRS~1915+105 is quite a peculiar case: it started its outburst in 1992 and since then it appears as a persistent X-ray source in the sky.

From gravitational wave observations, so far we have detected about 90~events in which two stellar-mass black holes (or a stellar-mass black hole and a neutron star or two neutron stars) merged to form a heavier black hole; see Fig.~\ref{f-bhgwem}. With the current sensitivity of the LIGO and Virgo experiments, we can detect a new merger event every few days, but the detection rate will significantly increase with the next generation of gravitational wave observatories.    

\begin{figure}[t]
\centering
\includegraphics[width=0.95\textwidth]{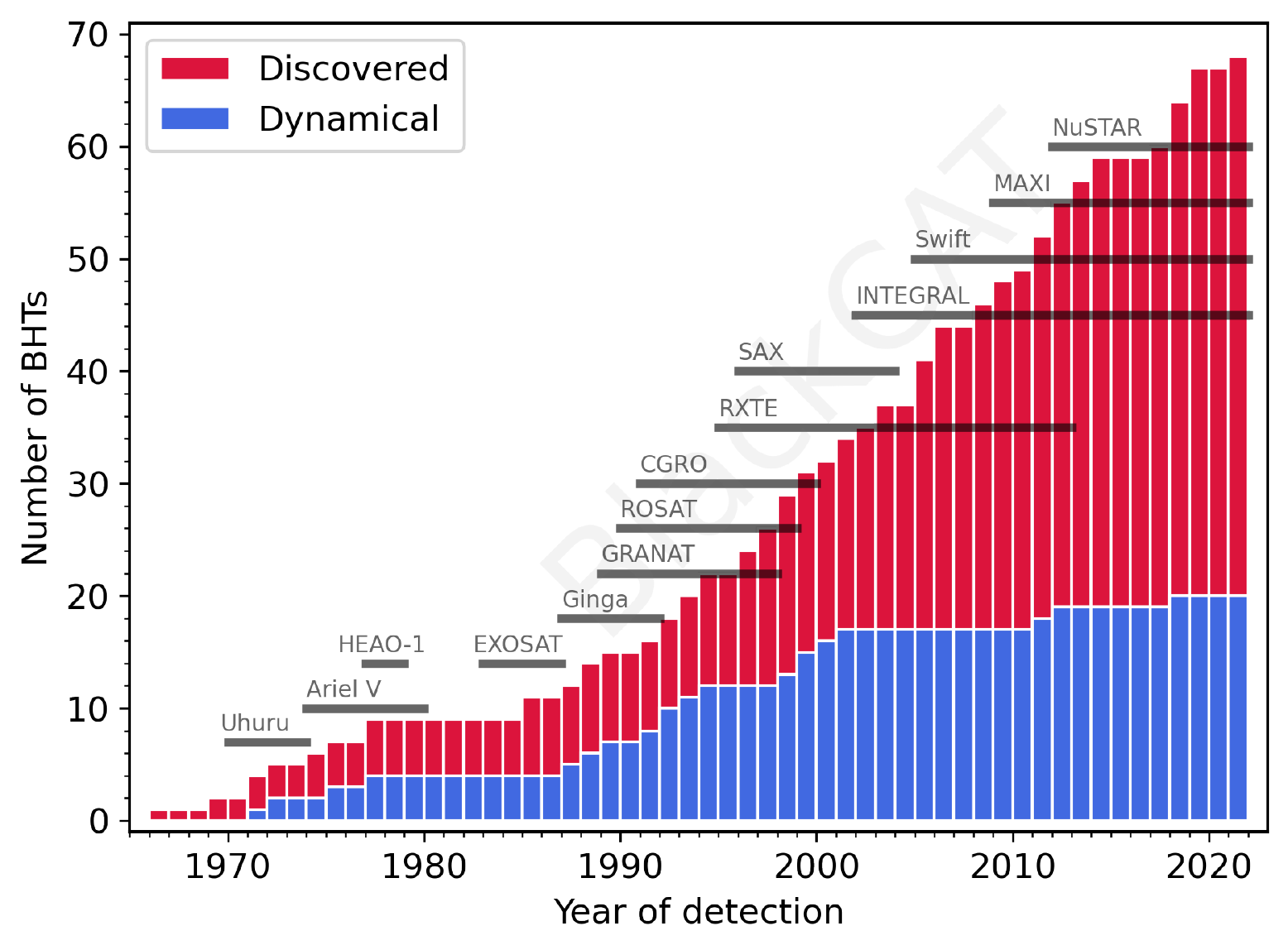}
\caption{Cumulative histogram of the number of discovered stellar-mass black holes in transient X-ray sources. Red bars are for new black holes and blue bars are for dynamically confirmed black holes. The horizontal gray bars show the main X-ray missions used to discover and study these black holes.
From the online BlackCAT catalog \url{https://www.astro.puc.cl/BlackCAT/} of Ref.~\cite{Corral-Santana:2015fud}.}
\label{f-bht}
\end{figure}

\begin{figure}[t]
\centering
\includegraphics[width=1.0\textwidth]{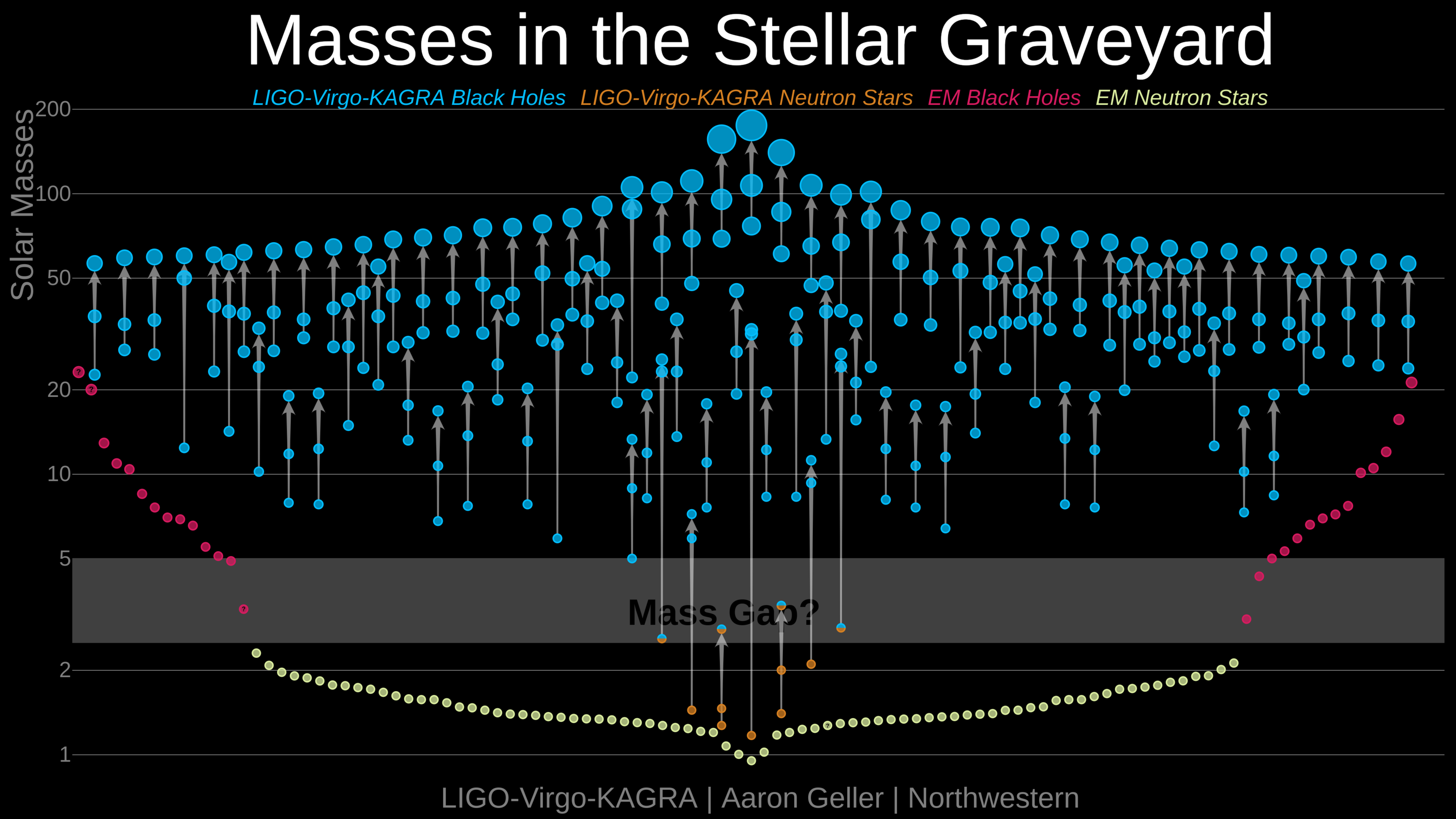}
\caption{Stellar-mass black holes and neutron stars with a robust mass measurement. Black holes (neutron stars) discovered with gravitational waves are in blue (orange) and black holes (neutron stars) in X-ray binaries are in magenta (green). Credit: LIGO-Virgo-KAGRA/Aaron Geller/Northwestern}
\label{f-bhgwem}
\end{figure}

\item {\it Supermassive black holes} are black holes with a mass in the range 10$^5$-10$^{10}$~$M_\odot$ and are found in galactic nuclei. Every middle-size or large galaxy seems to have a supermassive black hole at its center, while in the case of small galaxies the situation is more controversial: some small galaxies may have a supermassive black hole but other small galaxies may not.

While heavier objects can naturally migrate to the center of a multi-body system, and therefore it is not a surprise to find such supermassive objects in galactic nuclei, their exact origin is not completely understood: it is not like the case of stellar-mass black holes, which should be anyway expected as the final product of the evolution of very heavy starts. Supermassive black holes were certainly much smaller when they formed and have grown from merger with other black holes and accretion of the surrounding material. However, it is puzzling to observe black holes with masses of about 10$^{10}$~$M_\odot$ in high-redshift galaxies, when the Universe was only 1~Gyr old; see, e.g., Ref.~\cite{Wu2015}: a stellar-mass black hole formed from the first generation of stars would not have had the time to grow so much in such a short time without exceeding the Eddington limit. It is possible that the supermassive black holes we see today in galactic nuclei formed from the direct collapse of heavy clouds, and therefore their initial mass was a few order of magnitudes higher than the maximum mass of stellar-mass black holes~\cite{Volonteri:2010wz}. They may have also experienced some period of super-Eddington accretion and/or grown from the merger of several black holes~\cite{Volonteri:2010wz}. The actual mechanism is currently unknown, but it will be investigated by the next generation of gravitational wave observatories, which will have the capability of detecting black hole mergers at very high redshift, potentially even before the formation of the first stars.

\item {\it Intermediate-mass black holes} are black holes with a mass filling the gap between the stellar-mass and the supermassive black holes. However, unlike in the case of stellar-mass and supermassive black holes, so far there are no robust measurements of the masses of these objects, so technically we should speak about intermediate-mass black hole ``candidates''.

Some of these objects are in the so-called ultra-luminous X-ray sources, which are X-ray sources with a luminosity exceeding the Eddington limit of a stellar-mass black hole~\cite{Kaaret:2017tcn}. However, today we know that some ultra-luminous X-ray sources host neutron stars accreting above the Eddington limit~\cite{Bachetti:2014qsa}, so caution is necessary in the interpretation of these objects. Intermediate-mass black holes are expected to form at the center of stellar clusters from collision and merger of cluster members~\cite{Gebhardt:2005cy}. 

\end{itemize}

In this chapter, we review how the analysis of the X-ray radiation emitted from the inner part of accretion disks around black holes can test General Relativity in the strong field regime. The chapter is organized as follows. Section~\ref{sec:dc} introduces the astrophysical systems required for our tests. Section~\ref{sec:tests} shows how the analysis of X-ray spectra of accreting black holes can test General Relativity and which predictions and assumptions can be really studied. Section~\ref{sec:xrt} reviews the main techniques to test black holes with X-ray data. Section~\ref{sec:res} reviews the current observational constraints on the Kerr hypothesis. Section~\ref{sec:err} is devoted to discuss the systematic effects of these measurements and how robust these tests can be. Section~\ref{sec:c} is for the conclusions and future developments of this line of research.


\section{Disk-Corona Model}\label{sec:dc}

\begin{figure}[t]
\centering
\includegraphics[width=0.95\textwidth]{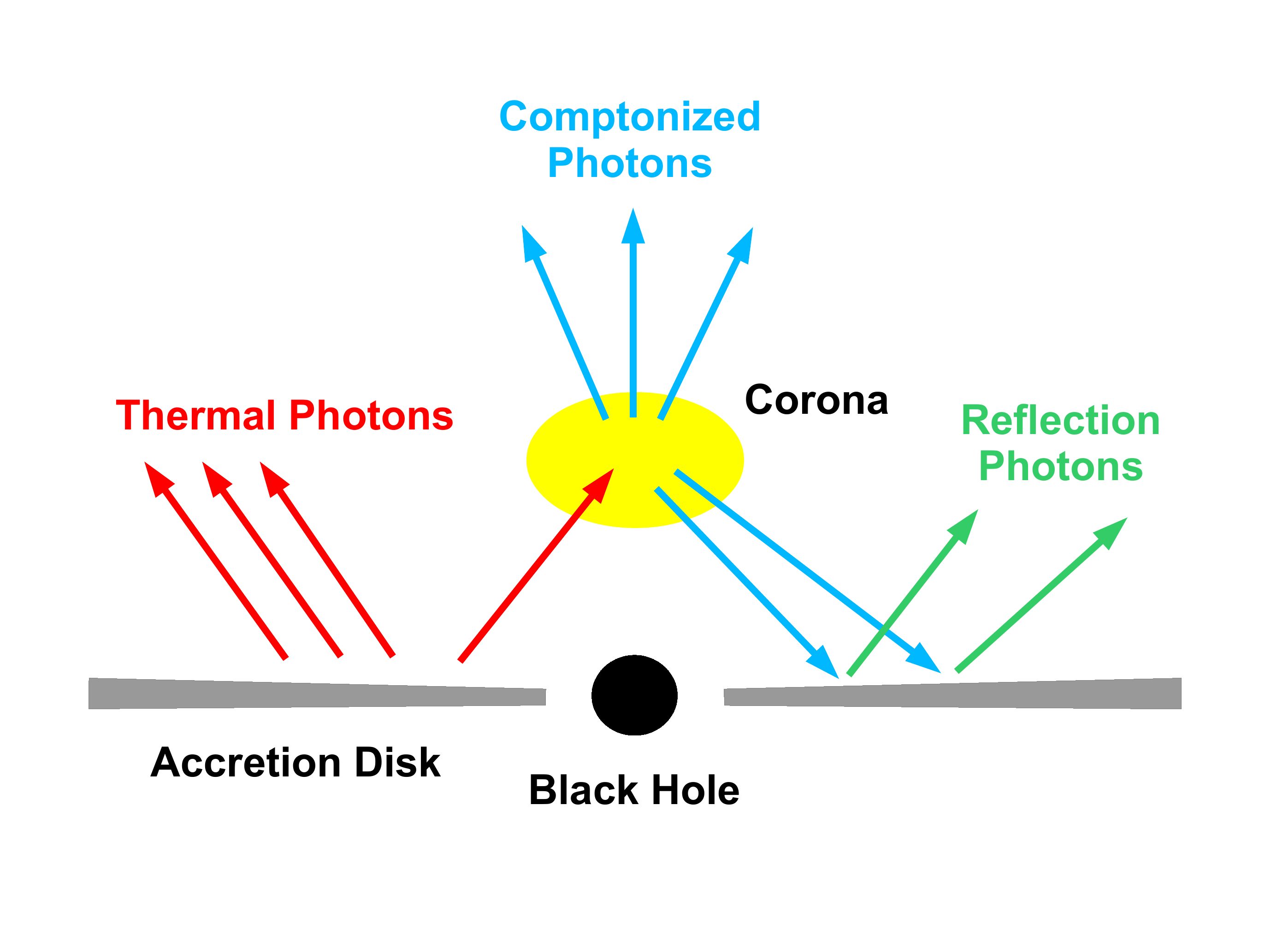}
\vspace{-0.8cm}
\caption{Disk-corona model. Figure from Ref.~\cite{Bambi:2021chr} under the terms of the Creative Commons Attribution 4.0 International License.}
\label{f-corona}
\end{figure}

The astrophysical system required for gravity tests with X-ray data is illustrated in Fig.~\ref{f-corona}, which is normally referred to as the disk-corona model~\cite{Bambi:2020jpe}. We have a black hole accreting from a geometrically thin and optically thick accretion disk. It is also useful to select sources in which the inner edge of the disk is at the innermost stable circular orbit (ISCO). The necessary conditions to have such an accretion disk are that the angular momentum of the gas is high and that the accretion luminosity is moderately high, say between about 5\% and 30\% of the Eddington luminosity of the source~\cite{Steiner:2010kd}. The high angular momentum of the accreting gas is necessary to have a disk. For lower accretion luminosities, the disk may be truncated at a radius larger than the ISCO. For higher accretion luminosities, the pressure of the gas becomes important and the disk may not be thin any longer near the black hole.

In the case of stellar-mass black holes in X-ray binaries, the angular momentum of the gas is always high enough to form an accretion disk, as the gas comes from the material transferring from the companion star to the black hole on the orbital plane. It is always necessary to check if the accretion disk is truncated or not, as this is a tricky point and it is not guaranteed that the inner edge is at the ISCO even if the accretion luminosity of the source is between 5\% and 30\% of the Eddington luminosity.

On the contrary, most supermassive black holes in galactic nuclei do not have a geometrically thin and optically thick accretion disk. For most supermassive black holes, including that at the center of the Milky Way, the angular momentum of the gas and the accretion rate are too low to have a thin disk. Active galactic nuclei (AGN) represent only about 0.035\% of all galactic nuclei and only a fraction of the supermassive black holes in AGN meet our requirements for gravity tests. Currently we do not know any intermediate-mass black hole candidate suitable for gravity tests.

In a geometrically thin and optically thick accretion disk, the gas is in local thermal equilibrium and every point of the surface of the disk has a blackbody-like spectrum. The whole disk has a multi-temperature blackbody-like spectrum, as the temperature of the gas increases as it falls through the gravitational well of the black hole. The temperature of the accretion disk roughly scales as $M^{-1/4}$, where $M$ is the black hole mass~\cite{Bambi:2017khi}. In the case of stellar-mass black holes, the thermal spectrum of the disk is peaked in the soft X-ray band (0.1-10~keV). In the case of supermassive black holes, the thermal spectrum is in the optical/UV band (1-100~eV).

\begin{figure}[t]
\centering
\includegraphics[width=0.95\textwidth]{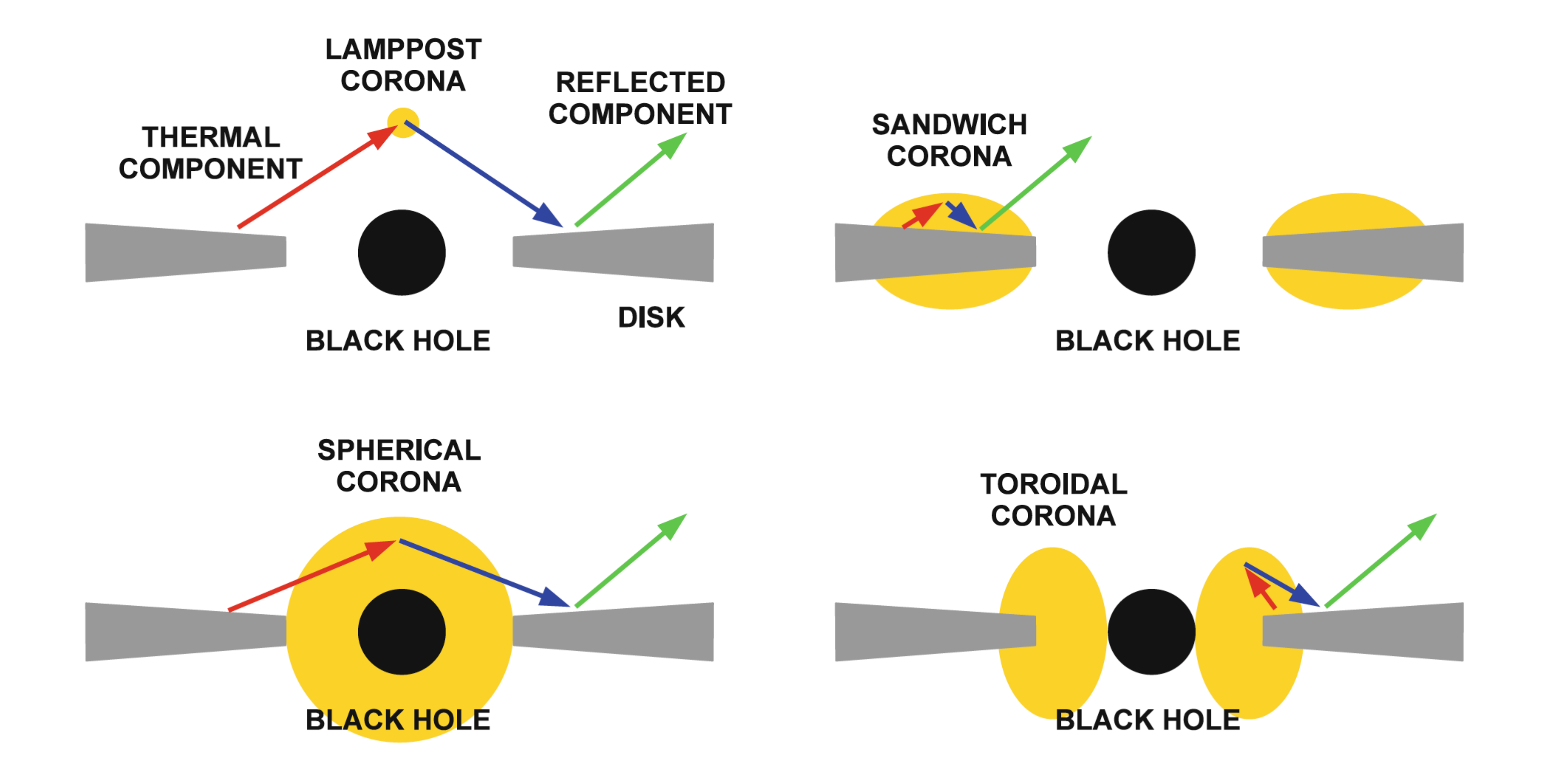}
\caption{Examples of possible coronal geometries: lamppost geometry (top left panel), sandwich geometry (top
right panel), spherical geometry (bottom left panel), and toroidal geometry (bottom right panel). Figure from Ref.~\cite{Bambi:2017khi}.}
\label{f-corona2}
\end{figure}

The {\it corona} is some hotter plasma (of the order of 100~keV) near the black hole and the inner part of the accretion disk, but its exact geometry is not yet well understood and it is expected to evolve in time. For example, the base of the jet, the atmosphere above the accretion disk, and/or the accretion flow in the plunging region between the inner edge of the disk and the black hole may act as coronae, and two or more coronae may coexist at the same time (see Fig.~\ref{f-corona2})~\cite{Bambi:2020jpe}.

Thermal photons from the disk can inverse Compton scatter off free electrons in the corona. The spectrum of the Comptonized photons is approximately described by a power law with a high-energy cutoff (and even a low-energy cutoff if we have low energy data). A fraction of the Comptonized photons can illuminate the disk: here we can have Compton scattering and absorption followed by fluorescent emission. The final result is a reflection spectrum of the disk, as shown in Fig.~\ref{f-corona} by the green arrows.

\begin{figure}[t]
\centering
\includegraphics[width=0.95\textwidth]{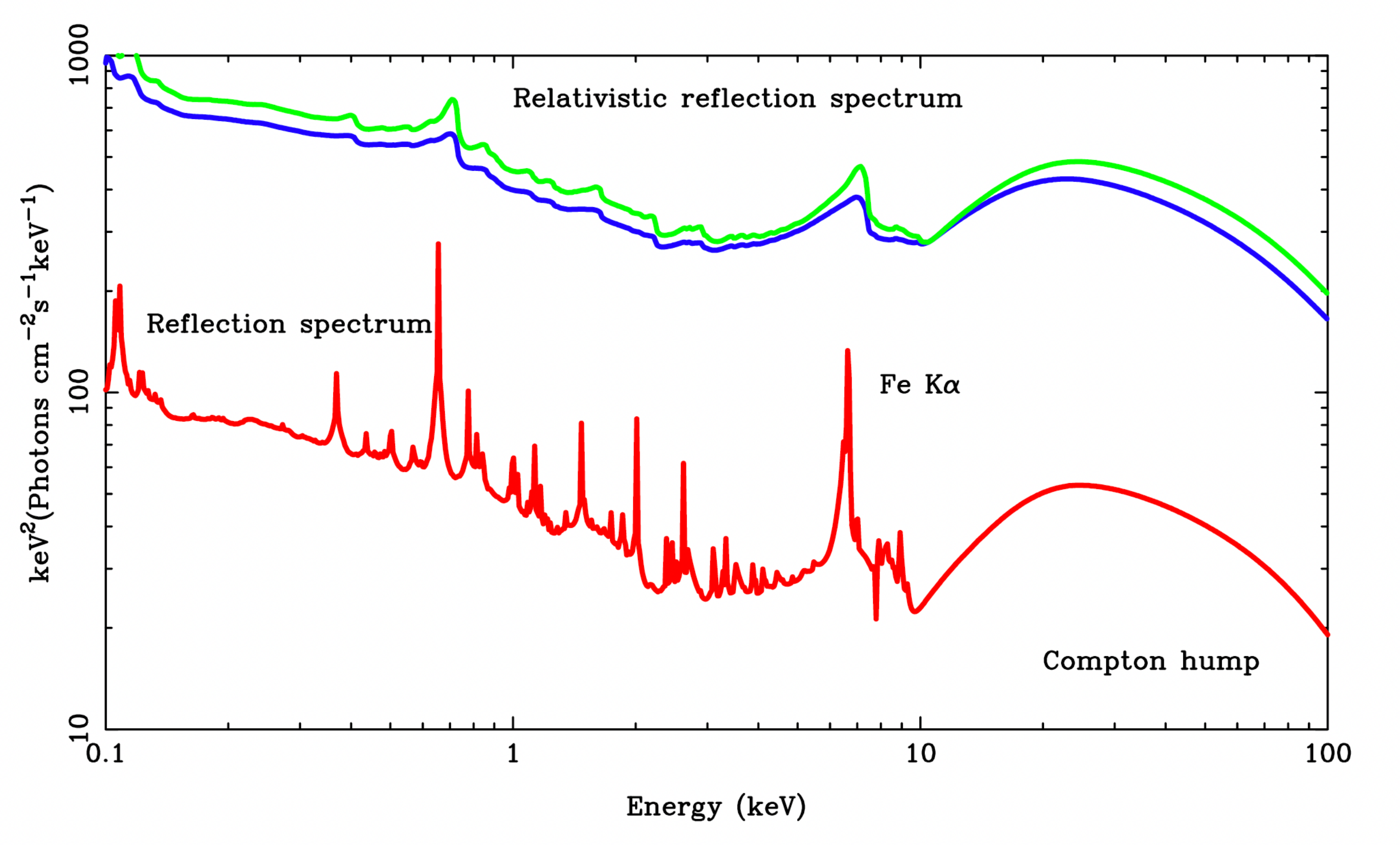}
\caption{Reflection spectrum in the rest-frame of the gas in the disk (red curve) and relativistic reflection spectra
of the whole disk as seen by a distant observer (green curve for a Schwarzschild spacetime and blue curve for a Kerr spacetime with spin parameter $a_* = 0.998$). Non-relativistic and relativistic reflection spectra have a different normalization to avoid overlapping. Figure from Ref.~\cite{Bambi:2020jpe}.}
\label{f-sp}
\end{figure}

In the rest-frame of the gas in the disk, the reflection spectrum is characterized by narrow fluorescent emission lines in the soft X-ray band and a Compton hump peaking at 20-30~keV (red spectrum in Fig.~\ref{f-sp})~\cite{Ross:2005dm,Garcia:2010iz}. The most prominent emission line is usually the iron K$\alpha$ complex, which is at 6.4~keV in the case of neutral or weakly ionized iron atoms and shifts up to 6.97~keV in the case of H-like iron ions (and there is no emission line in the case of fully ionized iron ions). The reflection spectrum of the whole disk as seen by a distant observer appears blurred as a result of relativistic effects (gravitational redshift and Doppler boosting) in the strong gravitational field around the black hole (green and blue spectra in Fig.~\ref{f-sp})~\cite{Fabian:1989ej,Laor:1991nc,Bambi:2017khi}.

\section{What can we test with X-ray data?}\label{sec:tests}

In general, the output of an X-ray detector is the measurement of the total photon count per spectral channel $C(h)$, where $h$ is the spectral channel in some engineering units. $C(h)$ can be written as 
\be\label{eq-ch}
C(h) = \tau \left[ \int dE \cdot R(h,E) \cdot A(E) \cdot s(E) \;  + \; B(h) \right] \; ,
\ee
where $\tau$ is the exposure time of the observation, $R(h,E)$ is the redistribution matrix, $A(E)$ is the effective area of the detector, $s(E)$ is the intrinsic spectrum of the source that we want to measure, and $B(h)$ is the instrumental background contribution.

\begin{figure}[t]
\centering
\includegraphics[width=0.95\textwidth,trim=2.0cm 2.5cm 2.0cm 2.0cm,clip]{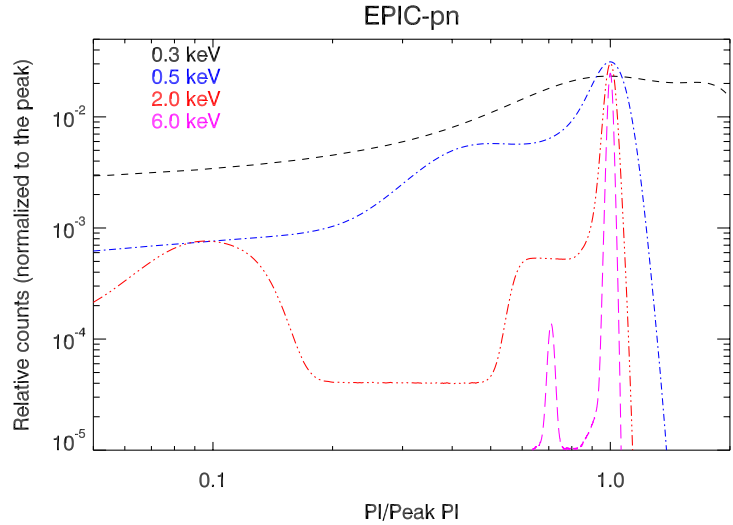}
\caption{Response of the EPIC-pn detector on board of \textsl{XMM-Newton} to monochromatic lines of 0.3, 0.5, 2.0, and 6.0~keV. PI (= pulse invariant) is an engineering unit for the spectral channel. Figure courtesy of Matteo Guainazzi.}
\label{f-rmf}
\end{figure}

The redistribution matrix $R(h,E)$ is related to the response of the detector and roughly corresponds to the probability that a photon of energy $E$ is detected in the channel $h$. In an ideal detector, the redistribution matrix should be a $\delta$-function: $R(h,E) = \delta(h - h_E)$. In a real detector, this is not possible and the width of the curve of $R(h,E)$ as a function of $h$ defines the instrumental resolution at the energy $E$. Fig.~\ref{f-rmf} shows the redistribution matrix of the EPIC-pn camera on board of \textsl{XMM-Newton} for four different photon energies. The effective area $A(E)$ depends on the optics, possible filters, and the detector, and roughly corresponds to the efficiency of these elements as a function of the photon energy. The shape of the measured spectrum, $C(h)$, is normally very different from the intrinsic spectrum of the source, $s(E)$, and is instead mainly determined by the characteristics of the instrument, so by $R(h,E)$ and $A(E)$.

In general, it is not possible to invert Eq.~(\ref{eq-ch}) and write $s(E)$ in terms of $C(h)$ because of cross-correlations among different energies. The standard strategy is to follow the so-called forward-folding approach, which is based on the following steps: 1)~we consider a theoretical model for the intrinsic spectrum of the source $s(E)$; 2)~we convolve the theoretical model with the instrument response to get the expected $C(h)$ for certain input parameters of the theoretical model; 3)~we compare the measured $C(h)$ with the expected $C(h)$ with some goodness-of-fit statistical test; 4)~we find the best-fit and we calculate the confidence intervals on the best-fit parameters.

The construction and the development of sufficiently sophisticated theoretical models to describe the intrinsic spectrum of the source $s(E)$ is crucial to get accurate measurements of the physical system. In the case of tests of General Relativity with black holes accreting from geometrically thin and optically thick accretion disks, we can expect to be able to test fundamental physics from the analysis of the thermal spectrum and of the reflection features. The Comptonized spectrum from the corona has a very simple shape -- usually it can be approximated by a power law spectrum with a high-energy cutoff -- and cannot be easily related to the physics occurring in the strong gravitational field of the black hole without knowing its exact geometry and location.

Theoretical models for the predictions of the thermal and reflection components are based on some assumptions, which can be grouped into two classes (see Tab.~\ref{t-h}): assumptions related to the underlying fundamental physics and those related to the astrophysical model. In the former group, we have the particle motion (e.g., motion of the particles of the accretion disk and motion of X-ray photons from the emission point on the disk to the detection point in the flat faraway region) and, for the calculation of the reflection component, the atomic physics in strong gravitational fields. For the astrophysical set-up, we need a model for the accretion disk and, for the calculation of the reflection component, even a model for the corona.

\begin{table}
\caption{Required ingredients for theoretical models to predict thermal and reflection spectra of thin disks around black holes} \label{t-h}
\centering
{\renewcommand{\arraystretch}{1.8}
\begin{tabular}{|c|}
\hline\hline
\hspace{1.0cm} Fundamental Physics \hspace{1.0cm} \\
\hline
Particle Motion \\
Atomic Physics (only for reflection) \\
\hline\hline
\end{tabular}
} \hspace{0.5cm}
{\renewcommand{\arraystretch}{1.8}
\begin{tabular}{|c|}
\hline\hline
\hspace{1.0cm} Astrophysical Model \hspace{1.0cm} \\
\hline
Accretion Disk \\
Corona (only for reflection) \\
\hline\hline
\end{tabular}
}
\end{table}

In General Relativity, the particle motion is the geodesic motion in the Kerr spacetime and the atomic physics is the same as in our laboratories on Earth (because in General Relativity the non-gravitational laws of physics reduce to those of Special Relativity in any locally inertial reference frame). However, in theories beyond General Relativity, the spacetime of a black hole may not be described by the Kerr solution, there may be deviations from geodesic motion (for all particle species or only for some particle species), and the atomic physics near a black hole may be different from that in our laboratories on Earth (e.g., some fundamental constants, like the fine structure constant $\alpha$ or the electron mass $m_e$, may be determined by the vacuum expectation value of a scalar field, which can be different around black holes and on Earth, and this would certainly have an impact on the energy levels of atoms and Compton scattering). If the astrophysical model is under control, we can think of relaxing some assumptions of the underlying fundamental physics and test these assumptions with X-ray observations of black holes.

Generally speaking, there are two possible approaches to test General Relativity:
\begin{itemize}
\item The {\it top-down} approach is the most natural and logical one: we want to compare General Relativity with some other theory of gravity. In such a case, we have to construct a theoretical model for General Relativity and another theoretical model for the other theory of gravity, fit the data with the two models, and check which of them can describe the data better and if it is possible to rule out the other one. There are at least two issues if we follow this approach. First, there are many theories of gravity beyond General Relativity and we should repeat the test for each of them. Second, we should be able to construct the theoretical model for the other theory of gravity and this is not so easy in general. For example, black hole solutions in theories beyond General Relativity are more often known only in the non-rotating limit while the complete rotating solution is unknown simply because it is too difficult to solve the field equations of the theory imposing only that the spacetime is stationary and axisymmetric. Without the complete rotating black hole solution, we cannot test the theory because astrophysical bodies have naturally a non-vanishing spin angular momentum and the spin of the black hole typically plays an important role in the shape of its X-ray spectrum.  
\item The {\it bottom-up} approach is an alternative strategy in which we want to test a specific assumption or prediction of General Relativity, which may be violated in other theories of gravity but we do not assume any of them in particular. In such a case, we have to employ a theoretical framework suitable to quantify possible violations of that assumption or prediction. Normally we have to introduce one or more parameters to do it. For example, we may be in the situation in which General Relativity is recovered if the new parameter vanishes and we have a violation of that assumption or prediction if the new parameter does not vanish. When we fit the data with the theoretical model and we measure the value of the new parameter, we can see if its value is consistent with its value in General Relativity. 
\end{itemize}

So far, most of the studies reported in the literature (with X-ray data but even with other electromagnetic observations) have focused on testing the Kerr hypothesis, namely that the spacetime around astrophysical black holes is described by the Kerr solution~\cite{Bambi:2015kza}. These tests have employed either known non-Kerr black hole solutions of particular theories of gravity or parametric black hole spacetimes in which deviations from the Kerr solution are quantified by phenomenological {\it deformation parameters} in the spirit of the bottom-up approach. In part, this choice is motivated by the fact that it is easier to test the Kerr hypothesis than geodesic motion and atomic physics. However, it is also because theories of gravity violating the Kerr hypothesis but preserving geodesic motion and atomic physics are the most conservative extensions of General Relativity: the field equations of the theory are different and uncharged black holes are not described by the Kerr solution\footnote{We note that a violation of the Kerr hypothesis implies a breakdown of General Relativity, but the opposite is not necessarily true; that is, there are theories of gravity in which uncharged black holes are still described by the Kerr metric~\cite{Psaltis:2007cw}.}, but gravity still universally and minimally couples to matter, so we have geodesic motion and the atomic physics around black holes is the same as in our laboratories on Earth.

Among the parametric black hole spacetimes proposed in the literature, X-ray tests of the Kerr hypothesis have widely used the Johannsen metric~\cite{Johannsen:2013szh}. Its line element in Boyer-Lindquist coordinates is
\be\label{eq-j}
ds^2 &=&-\frac{{\Sigma}\left(\Delta-a^2A_2^2\sin^2\theta\right)}{B^2}dt^2
+\frac{{\Sigma}}{\Delta A_5}dr^2+{\Sigma} d\theta^2 \nonumber\\
&& +\frac{\left[\left(r^2+a^2\right)^2A_1^2-a^2\Delta\sin^2\theta\right]{\Sigma}\sin^2\theta}{B^2}d\phi^2 \nonumber\\
&& -\frac{2a\left[\left(r^2+a^2\right)A_1A_2-\Delta\right]{\Sigma}\sin^2\theta}{B^2}dtd\phi \, , 
\ee
where
\be
\Sigma &=& r^2 + a^2 \cos^2\theta \, , \\
\Delta &=& r^2 - 2 M r + a^2 \, , \\
B &=& \left(r^2+a^2\right)A_1-a^2A_2\sin^2\theta
\ee
and the functions $f$, $A_1$, $A_2$, and $A_5$ are defined as
\be
f &=& \sum^\infty_{n=3} \epsilon_n \frac{M^n}{r^{n-2}} \, , \qquad
A_1 = 1 + \sum^\infty_{n=3} \alpha_{1n} \left(\frac{M}{r}\right)^n \, , \\
A_2 &=& 1 + \sum^\infty_{n=2} \alpha_{2n} \left(\frac{M}{r}\right)^n \, , \qquad
A_5 = 1 + \sum^\infty_{n=2} \alpha_{5n} \left(\frac{M}{r}\right)^n 
\ee
This form of the Johannsen metric is consistent with the Newtonian limit and without constraints from Solar System experiments. If all deformation parameters vanish, we exactly recover the Kerr solution. The leading order deformation parameters are $\epsilon_3$, $\alpha_{13}$, $\alpha_{22}$, and $\alpha_{52}$. $\alpha_{13}$ is the deformation parameter with the strongest impact on the shape of the spectrum of thin disks, and thus the most widely studied.


\section{X-ray techniques}\label{sec:xrt}

The spectral analyses of the thermal component and of the reflection features of thin accretion disks are currently the two leading techniques to test General Relativity with black hole X-ray data and are reviewed in Subsections~\ref{subsec:cfm} and \ref{subsec:xrs}, respectively. Other X-ray techniques are listed in Subsection~\ref{subsec:other}, but none of them can currently be used to test General Relativity. For a review on electromagnetic techniques in general, see Ref.~\cite{Bambi:2015kza}.

\subsection{Continuum-fitting method}\label{subsec:cfm}

The spectral analysis of the thermal component of geometrically thin and optically thick accretion disks around black holes is normally referred to as the continuum-fitting method, because the thermal component is described by a continuum spectrum without features. This technique was originally proposed by Shuang-Nan Zhang in Ref.~\cite{Zhang:1997dy} to measure black hole spins (assuming General Relativity) and then developed by the Harvard group of Jeff McClintock and Ramesh Narayan (see, e.g., Refs.~\cite{McClintock:2013vwa,McClintock:2011zq}). So far the continuum-fitting method has provided a spin measurement for about 20~stellar-mass black holes~\cite{Draghis:2022ngm}. The continuum-fitting method is normally used only to estimate the spins of stellar-mass black holes in X-ray binary systems. In the case of supermassive black holes in AGN, the thermal spectrum of the disk is peaked in the UV band, where dust absorption does not permit an accurate measurement.

Diego Torres was the first to consider the continuum-fitting method as a technique to test fundamental physics~\cite{Torres:2002td}, followed by other authors that calculated the thermal spectra of thin accretion disks around non-Kerr black holes or other exotic compact objects like wormholes and boson stars~\cite{Pun:2008ae,Harko:2008vy,Harko:2009xf,Harko:2009rp,Harko:2010ua,Kovacs:2010xm,Bambi:2011jq,Bambi:2012tg}. The thermal spectrum of a thin disk has quite a simple shape: if we assume the Kerr metric, we can estimate the black hole spin parameter, but relaxing the Kerr hypothesis there is normally a degeneracy between the spin parameter of the object and possible deviations from the Kerr solution~\cite{Kong:2014wha}.

In Ref.~\cite{Zhou:2019fcg}, we presented {\tt nkbb}\footnote{{\tt nkbb} is public and available at \url{https://github.com/ABHModels}.}, which is the first -- and currently the only -- thermal model suitable to analyze X-ray data. The first application of {\tt nkbb} was the analysis of \textsl{RXTE} spectra of the stellar-mass black hole in LMC~X-1~\cite{Tripathi:2020qco} and we confirmed the strong degeneracy between the black hole spin parameter and possible deviations from the Kerr solution.

\begin{figure}[t]
\centering
\includegraphics[width=0.49\textwidth,trim=0.7cm 0.0cm 1.8cm 0.0cm,clip]{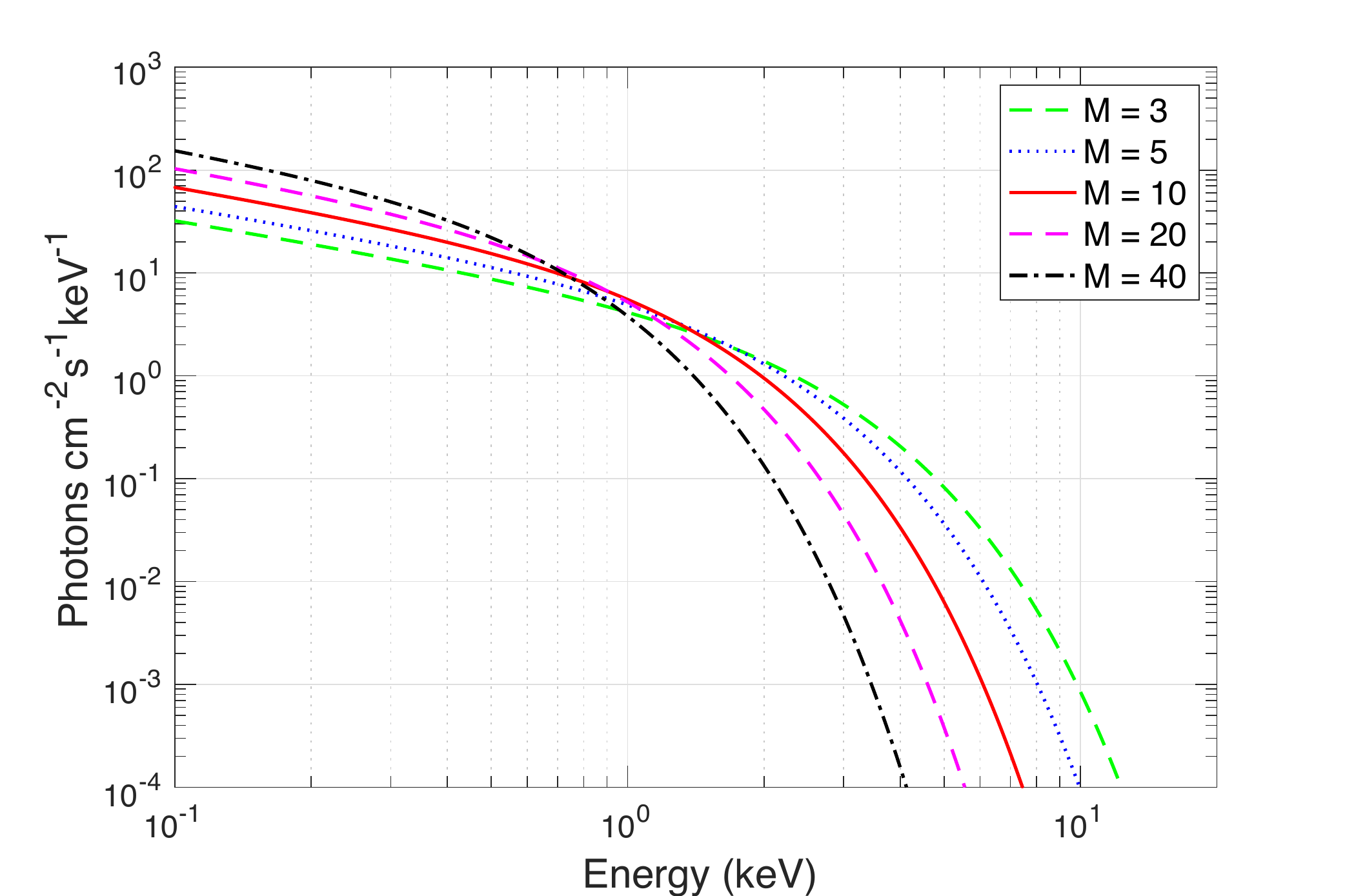}
\includegraphics[width=0.49\textwidth,trim=0.7cm 0.0cm 1.8cm 0.0cm,clip]{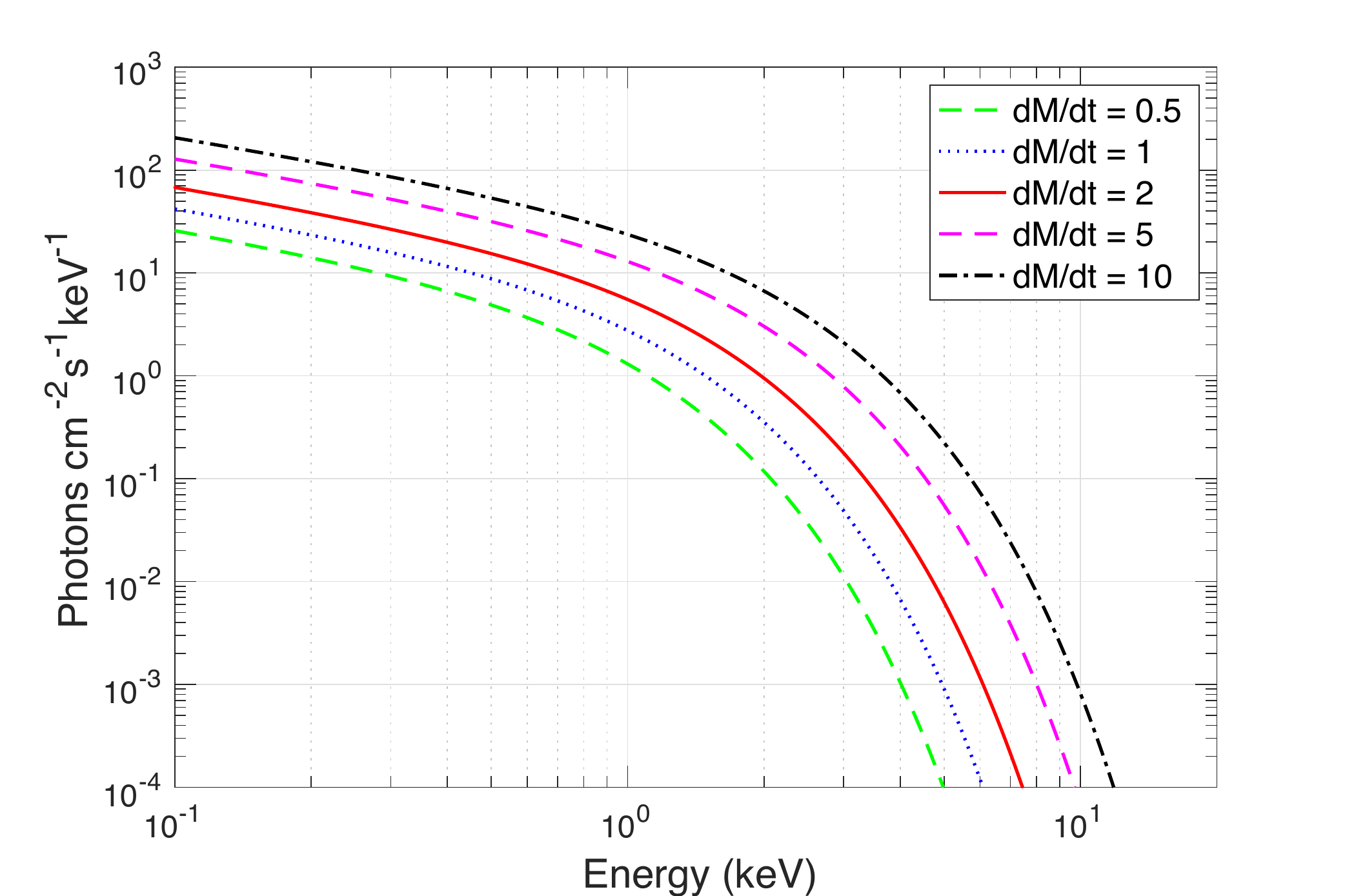}
\includegraphics[width=0.49\textwidth,trim=0.7cm 0.0cm 1.8cm 0.0cm,clip]{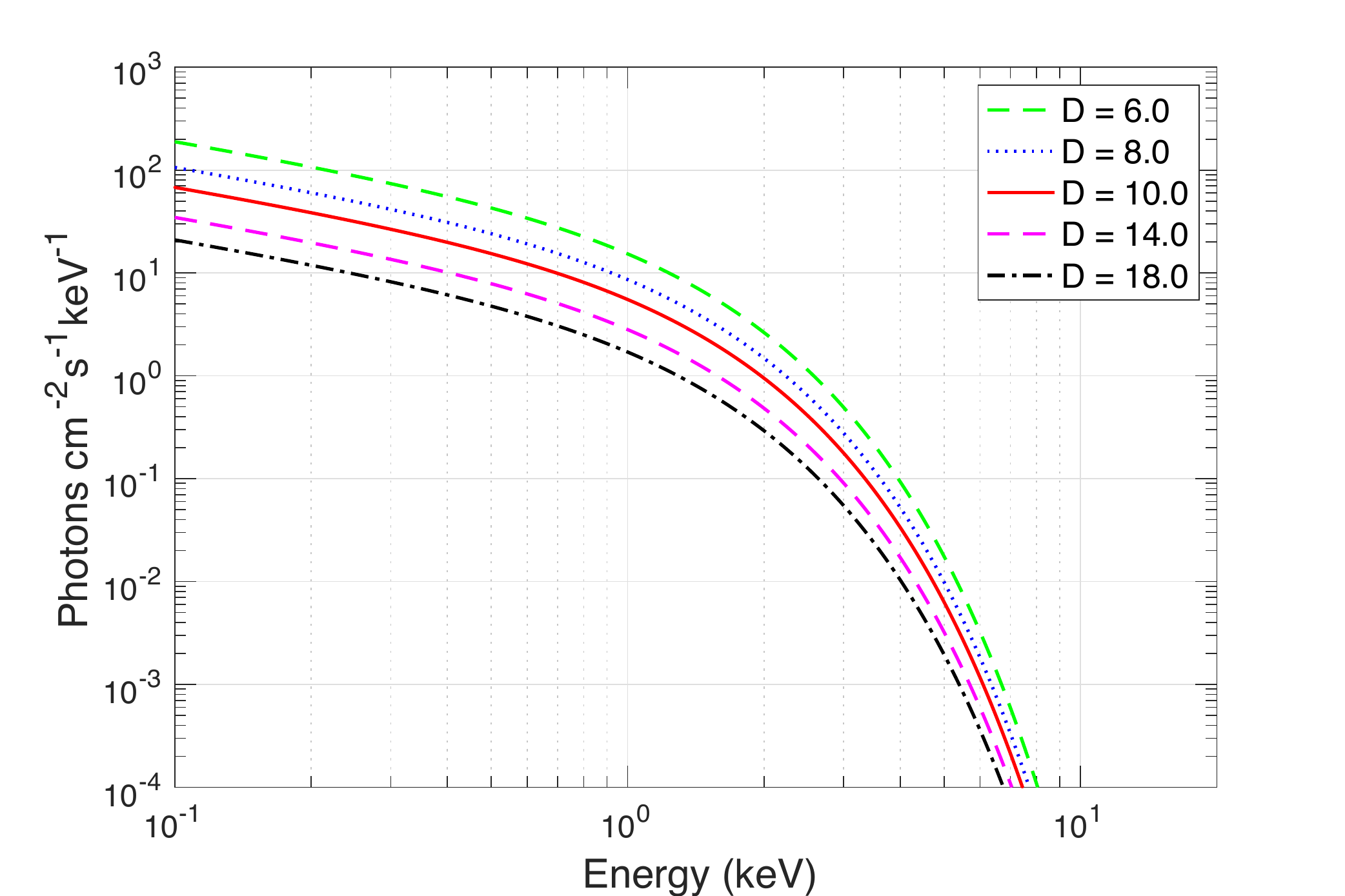}
\includegraphics[width=0.49\textwidth,trim=0.7cm 0.0cm 1.8cm 0.0cm,clip]{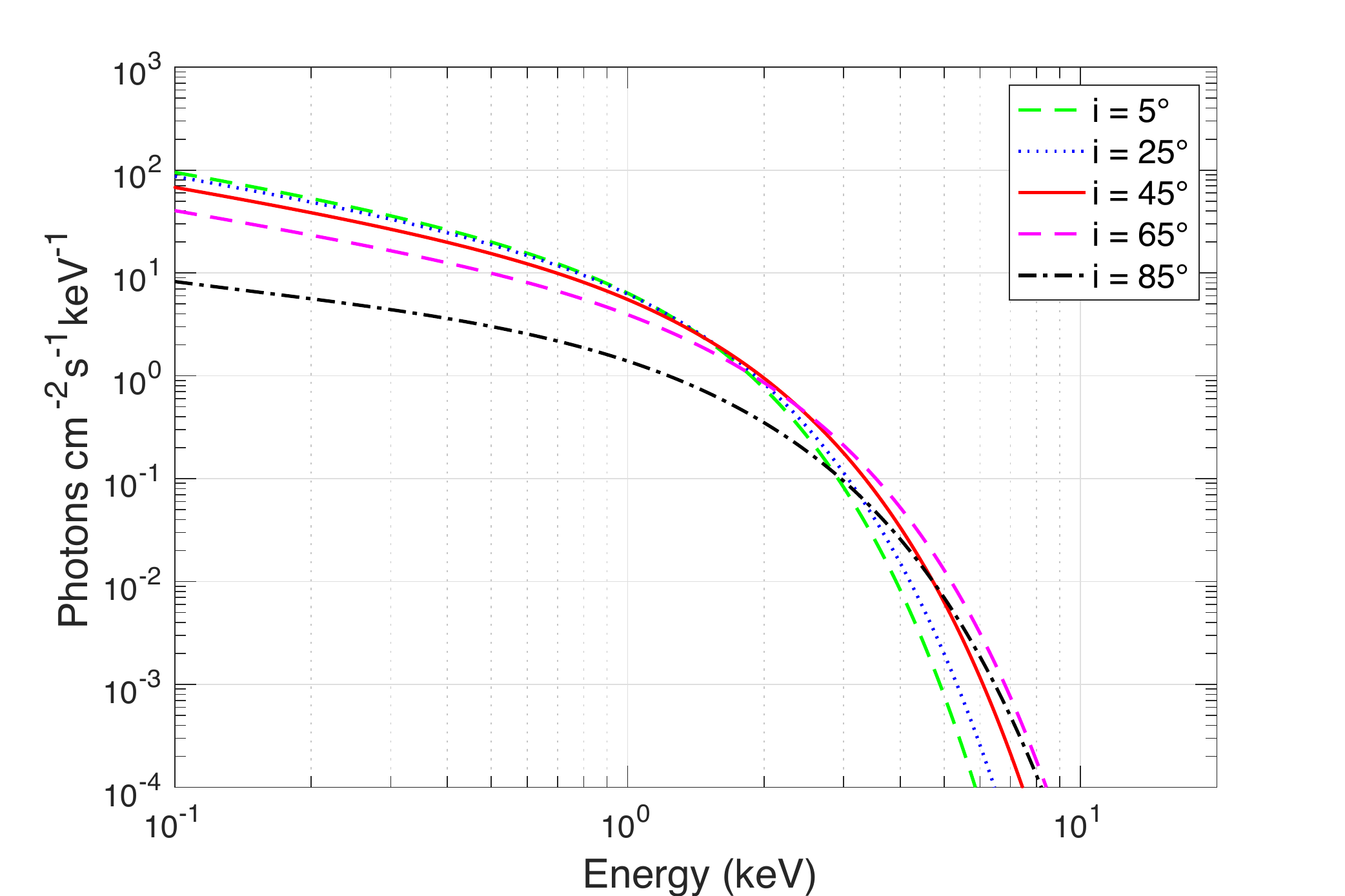}
\includegraphics[width=0.49\textwidth,trim=0.7cm 0.0cm 1.8cm 0.0cm,clip]{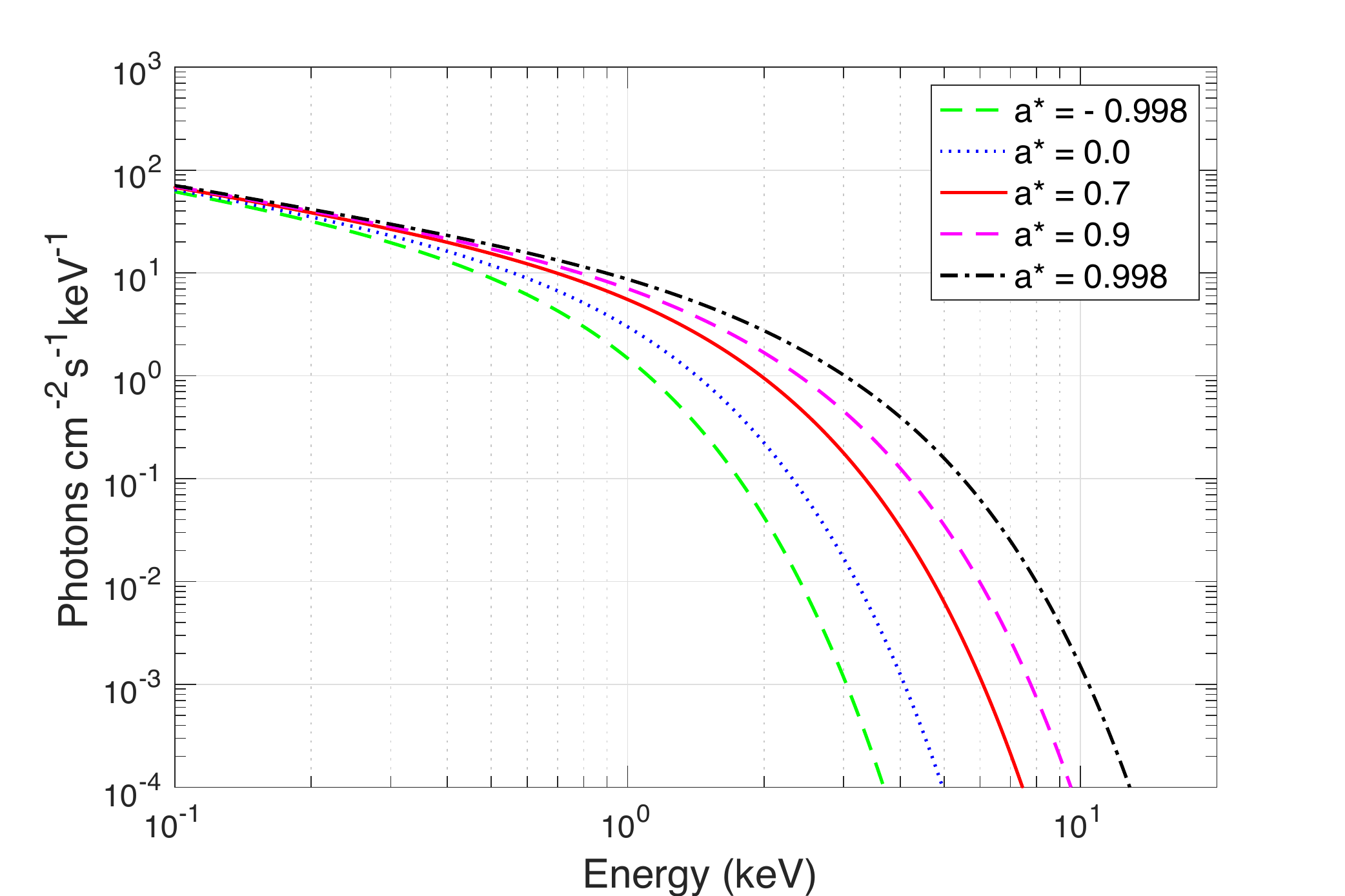}
\includegraphics[width=0.49\textwidth,trim=0.7cm 0.0cm 1.8cm 0.0cm,clip]{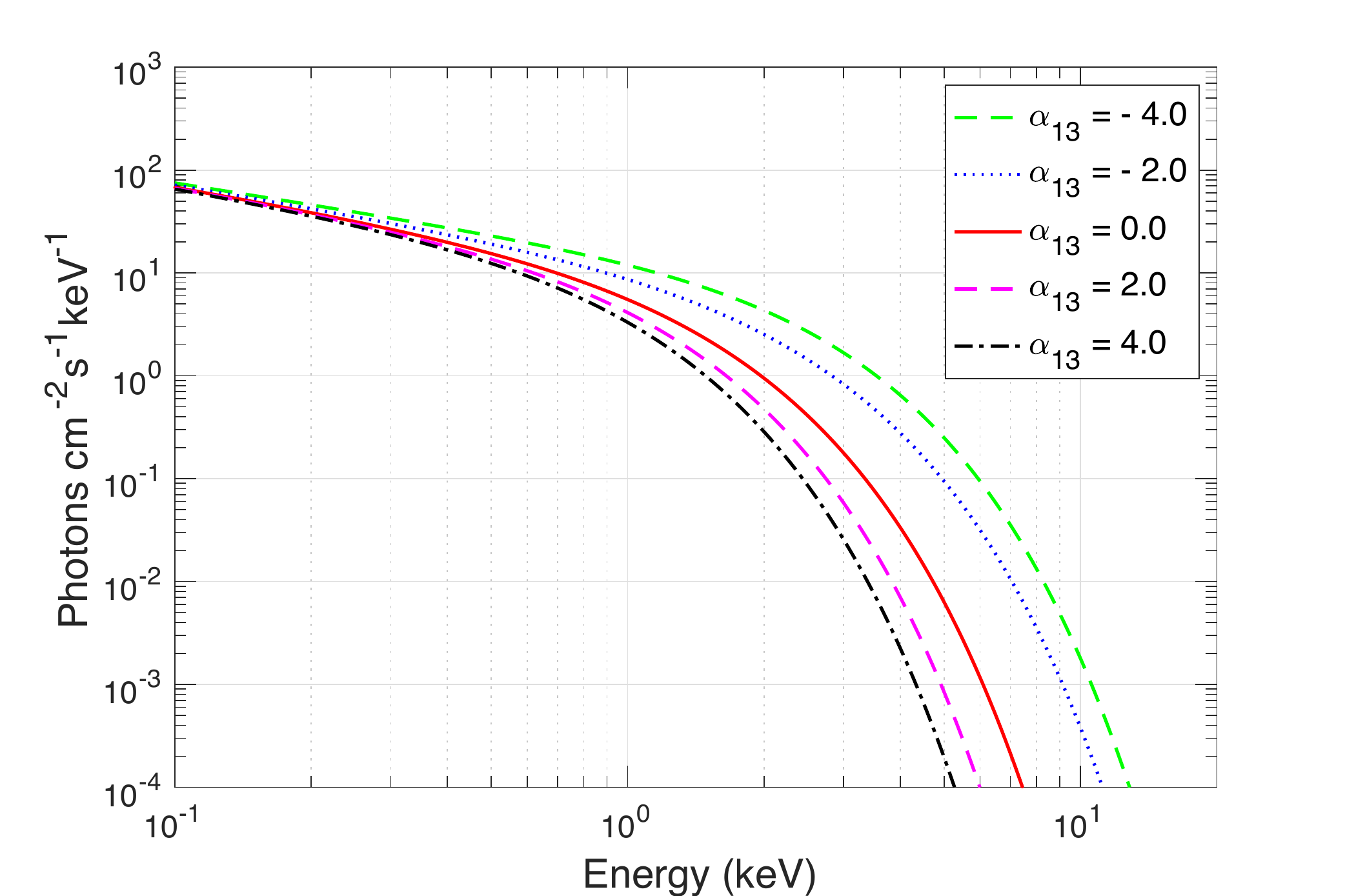}
\caption{Synthetic thermal spectra of thin disks in the Johannsen spacetime as calculated by {\tt nkbb} for different values of the model parameters. $M$ in $M_\odot$, $\dot{M}$ in $10^{18}$~g/s, and $D$ in kpc. When not shown, the values of the model parameters are: $M = 10 \, M_\odot$, $\dot{M} = 2 \cdot 10^{18}$~g/s, $D = 10$~kpc, $i = 45^\circ$, $a_* = 0.7$, and $\alpha_{13} = 0$. Figure from Ref.~\cite{Zhou:2019fcg}.}
\label{f-nkbb}
\end{figure}

If we assume geodesic motion, the calculation of the thermal spectrum of a thin disk requires two inputs: the spacetime metric and an accretion disk model. The standard framework for the description of geometrically thin and optically thick disks is the Novikov-Thorne model~\cite{Novikov:1973kta,Page:1974he}. The disk is on the equatorial plane, perpendicular to the spin axis of the central object. The material of the disk follows nearly-geodesic circular orbits on the equatorial plane. The inner edge of the disk is at the ISCO and the material that reaches the ISCO quickly plunges onto the black hole. From the conservation of mass, energy, and angular momentum, we can derive the time-averaged radial structure of the accretion disk. The calculation of thermal spectra of Novikov-Thorne disks have been extensively discussed in the literature in Kerr and non-Kerr spacetimes~\cite{Li:2004aq,Bambi:2011jq,Bambi:2012tg,Bambi:2017khi}. In the Kerr spacetime, eventually the thermal spectrum of the disk depends on 5 parameters: the black hole mass $M$, the black hole mass accretion rate $\dot{M}$, the black hole distance $D$, the inclination angle of the disk with respect to the line of sight of the distant observer $i$, and the black hole spin parameter $a_*$. If we employ a non-Kerr spacetime with a deformation parameter to quantify possible deviations from the Kerr solution, the model will depend also on this sixth parameter. Fig.~\ref{f-nkbb} shows thermal spectra of thin disks in the Johannsen spacetime as calculated by {\tt nkbb} for different values of the model parameters.

\begin{figure}[t]
\centering
\includegraphics[width=0.95\textwidth,trim=1.0cm 2.5cm 0.0cm 2.0cm,clip]{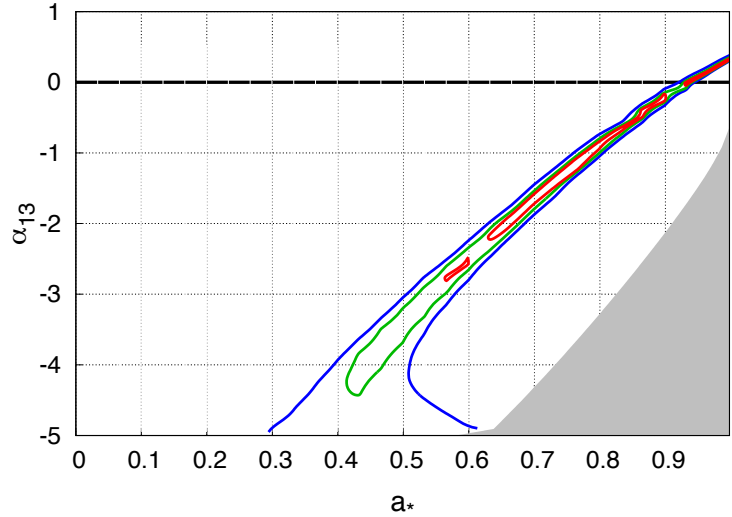}
\caption{Constraints on the spin parameter $a_*$ and the Johannsen deformation parameter $\alpha_{13}$ from the analysis of the thermal spectrum of the stellar-mass black hole in LMC~X-1 with \textsl{RXTE} data reported in Ref.~\cite{Tripathi:2020qco}. The red, green, and blue curves represent, respectively, the 68\%, 90\% and 99\% confidence level curves for two relevant parameters ($\Delta\chi^2 = 2.30$, 4.61, and 9.21, respectively). The horizontal black thick line at $\alpha_{13} = 0$ marks the Kerr solution. The gray region is ignored in the analysis because includes pathological spacetimes. Figure from Ref.~\cite{Tripathi:2020qco}.}
\label{f-lmcx1}
\end{figure}

In general, there is a strong degeneracy among these parameters. The continuum-fitting method requires to have independent estimate of the black hole mass $M$, the black hole distant $D$, and the inclination angle of the disk $i$ (for example, from optical observations of the companion stars), and one can fit the data to estimate the black hole spin parameter $a_*$, the black hole mass accretion rate $\dot{M}$, and a possible deformation parameter. If we fix $M$, $D$, $i$, and $\dot{M}$, the shape of the spectrum is mainly determined by the location of the inner edge of the disk, which is supposed to be at the ISCO but this requires an accurate selection of the spectra to analyze~\cite{Steiner:2010kd}. More specifically, the high-energy cutoff of the spectrum is mainly determined by the radiative efficiency at the ISCO radius, $\eta = 1 - E_{\rm ISCO}$~\cite{Kong:2014wha}, where $E_{\rm ISCO}$ is the specific energy of a test-particle at the ISCO radius and only depends on the metric coefficients $g_{tt}$, $g_{t\phi}$, and $g_{\phi\phi}$~\cite{Bambi:2017khi}. In the Kerr spacetime, $E_{\rm ISCO}$ is determined by the black hole spin parameter $a_*$ and it is relatively straightforward to measure $a_*$ from the analysis of the thermal spectrum of the disk. In non-Kerr spacetimes, it is not so easy. If the deformation parameter appears in the metric coefficients $g_{tt}$, $g_{t\phi}$, and $g_{\phi\phi}$, there is normally a strong degeneracy between $a_*$ and the deformation parameter, and we find the kind of constraints shown in Fig.~\ref{f-lmcx1}. If the deformation parameter appears only in the metric coefficients $g_{rr}$ and/or $g_{\theta\theta}$, its impact on the shape of thermal spectra is normally too weak to be measured.

\subsection{X-ray reflection spectroscopy}\label{subsec:xrs}

X-ray reflection spectroscopy refers to the spectral analysis of the reflection component. This technique is sometimes called the iron line method, especially in the old literature, as the iron K$\alpha$ line is normally the most prominent fluorescent emission line of the reflection spectrum and early studies, based on observations of poor quality, focused only on the analysis of the iron K$\alpha$ line. Andy Fabian was the first to propose to use the relativistically blurred reflection lines to probe the strong gravity region around black holes~\cite{Fabian:1989ej}, first detected in Cygnus~X-1 and reported in Ref.~\cite{barr1985}. The first clear detection of a broadened iron line was reported in Ref.~\cite{Tanaka:1995en} and observed in the spectrum of the supermassive black hole in MCG--06--30--15. Starting from the work by Laura Brenneman and Chris Reynolds in Ref.~\cite{Brenneman:2006hw}, X-ray reflection spectroscopy has been mainly developed for measuring black hole spins and today we have over 40~stellar-mass black holes and over 40~supermassive black holes with a spin measurement reported in the literature from the analysis of the reflection features of their spectra~\cite{Bambi:2020jpe,Draghis:2022ngm}.

The first study of a relativistically broadened iron line in a non-Kerr spacetime was reported by Youjun Lu and Diego Torres in Ref.~\cite{Lu:2002vm}, and later extended by other authors to other non-Kerr spacetimes~\cite{Schee:2008fc,Johannsen:2012ng,Bambi:2012at,Bambi:2013jda,Bambi:2013hza,Jiang:2015dla,Zhou:2016koy,Ni:2016rhz,Ni:2016uik,Cao:2016zbh,Shen:2016acv,Zhou:2017glv,Zhang:2017unx,Liu:2018bfx,Nampalliwar:2018iru,Zhang:2018xzj,Yang:2018wye}. In Refs.~\cite{Bambi:2016sac,Abdikamalov:2019yrr}, we presented the first -- and currently the only -- full reflection model (i.e. for the whole reflection spectrum, not only an iron line) suitable to analyze X-ray data and test the Kerr hypothesis. The model is called {\tt relxill\_nk}\footnote{{\tt relxill\_nk} is public and available at \url{https://github.com/ABHModels}.}, because it is an extension of the {\tt relxill} package developed by Thomas Dauser and Javier Garcia~\cite{Dauser:2013xv,Garcia:2013oma,Garcia:2013lxa}. The first test of the Kerr hypothesis from X-ray reflection spectroscopy was reported in Ref.~\cite{Cao:2017kdq}. In the past couple of years, {\tt relxill\_nk} has been developed to include the thickness of the disk~\cite{Abdikamalov:2020oci}, non-trivial ionization and density profiles~\cite{Abdikamalov:2021rty,Abdikamalov:2021ues}, and specific coronal geometries~\cite{Riaz:2020svt}.

The calculation of reflection spectra of geometrically thin and optically thick accretion disks in Kerr and non-Kerr spacetimes have been extensively discussed in the literature. There are two independent blocks. One of these two blocks is represented by the calculation of the reflection spectrum in the rest-frame of the gas of the disk: in General Relativity, the non-gravitational laws of physics reduce to those of Special Relativity in any locally inertial frame and therefore the calculation of the non-relativistic reflection spectrum only depends on atomic physics~\cite{Ross:2005dm,Garcia:2010iz}. The second block is represented by the relativistic effects that make the original reflection spectrum blurred, as photons emitted from different points of the disk have a different redshift/blueshift when they arrive at the detection point far from the source~\cite{Bambi:2016sac,Bambi:2017khi}. From the Novikov-Thorne model, the material in the disk follows nearly-geodesic equatorial circular orbits and there is no emission from the plunging region, between the inner edge of the accretion disk and the black hole. These are the only ingredients of the Novikov-Thorne model entering the calculations of the reflection spectrum. The inner edge of the disk can be at the ISCO, and it is useful to select sources with the inner edge at the ISCO if we want to measure the black hole spin parameter (assuming General Relativity) or test General Relativity, but this is not strictly necessary.

Theoretical models for reflection spectra of accretion disks have normally several parameters, which should be measured in the data analysis process. In the {\tt relxill\_nk} package, there are a number of ``flavors'', namely different versions of the model based on different assumptions. The default flavor has 16~parameters. The spacetime metric is described by the spin parameter of the central object $a_*$ and by a deformation parameter introduced to quantify possible deviations from the Kerr solution (for instance, the deformation parameter $\alpha_{13}$ of the Johannsen metric, but {\tt relxill\_nk} can easily work with any stationary, axisymmetric, and asymptotically-flat metric given in analytical form). The mass of the central object does not directly enter the calculation of the reflection spectrum. The accretion disk has an inner radius $R_{\rm in}$ (which can be set at the ISCO or at some larger value, but it can also be left free in the fit) and an outer radius $R_{\rm out}$. The spectrum of the corona illuminating the disk is described by a power law with a high-energy cutoff, and there are thus two parameters: the photon index $\Gamma$ and the high-energy cutoff $E_{\rm cut}$. For a corona with arbitrary geometry, the emissivity profile of the accretion disk $\varepsilon$ is modeled with a twice broken power law, namely there are three regions:
\be\label{eq-tbpl}
\varepsilon (r) = 
\begin{cases}
\left( \frac{1}{r} \right)^{q_1} & \text{if } r < R_{\rm br1} \; , \\
\left( \frac{R_{\rm br1}}{r} \right)^{q_2} \left( \frac{1}{R_{\rm br1}} \right)^{q_1} & \text{if } R_{\rm br1} < r < R_{\rm br2} \; , \\
\left( \frac{R_{\rm br2}}{r} \right)^{q_3} \left( \frac{R_{\rm br1}}{R_{\rm br2}} \right)^{q_2} \left( \frac{1}{R_{\rm br1}} \right)^{q_1} & \text{if } r > R_{\rm br2} \; .
\end{cases}
\ee
and therefore five parameters (three emissivity indices for the inner/central/outer regions, $q_1$/$q_2$/$q_3$, and two breaking radii separating the three regions, $R_{\rm br1}$ and $R_{\rm br2}$). The orientation of the accretion disk with respect to the distant observer is described by the inclination angle of the disk with respect to the line of sight of the observer, $i$. The material of the disk is characterized by the ionization parameter $\xi$ (which is constant over the whole disk in the default flavor) and the iron abundance $A_{\rm Fe}$. The redshift of the source $z$ can be set to 0 in the case of X-ray binaries, as the relative motion between us and the source is negligible, and to the value of the host galaxy measured by optical observations in the case of AGN. The last parameter is the normalization of the component, which is always determined by the fit during the data analysis process.

\begin{figure}[t]
\centering
\includegraphics[width=0.95\textwidth,trim=0.2cm 0.2cm 0cm 0cm,clip]{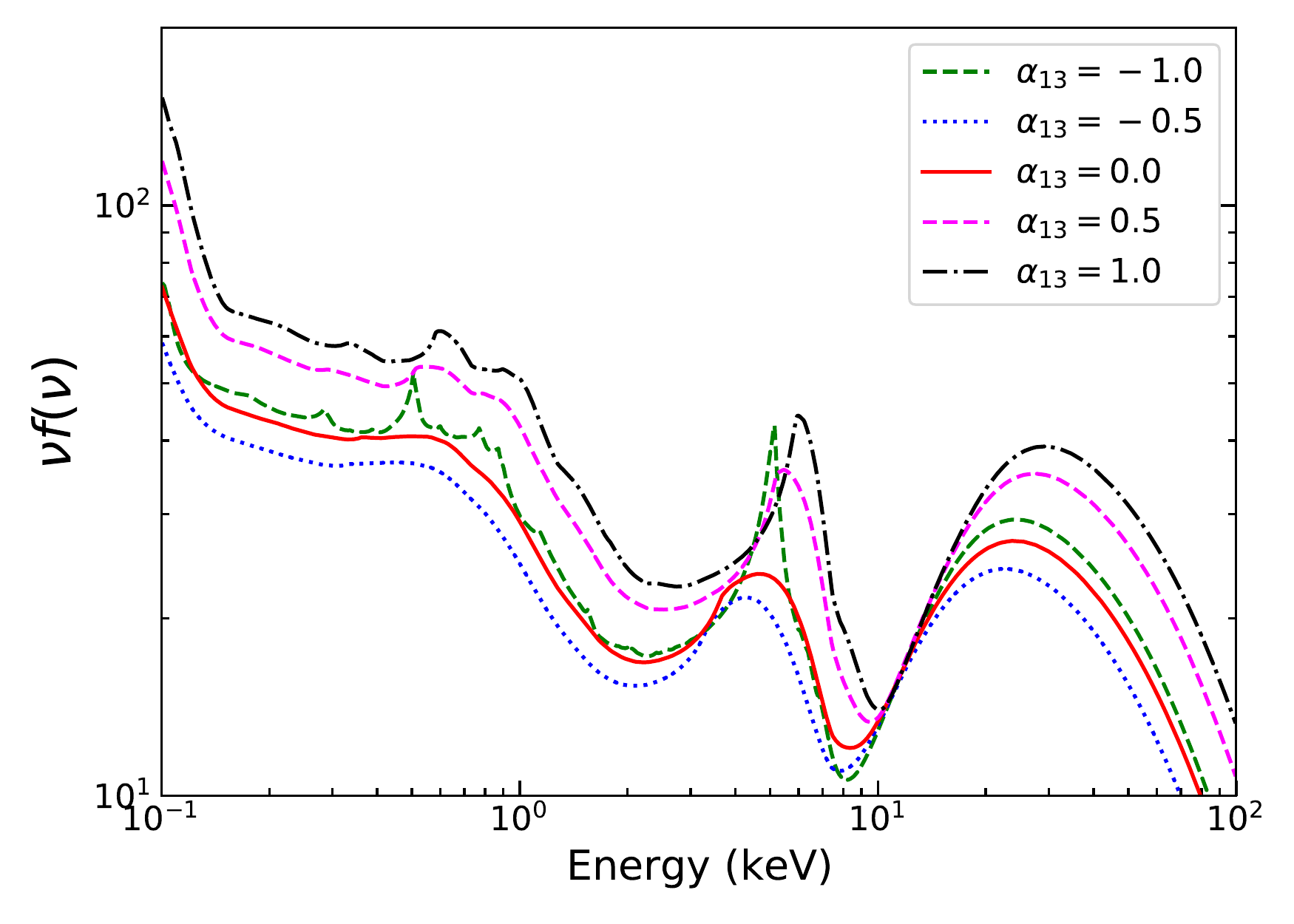}
\caption{Examples of synthetic reflection spectra of thin disks in the Johannsen spacetime as calculated by {\tt relxill\_nk} for different values of the Johannsen deformation parameter $\alpha_{13}$. The values of the other parameters of the models do not change. Figure from Ref.~\cite{Bambi:2021chr} under the terms of the Creative Commons Attribution 4.0 International License.}
\label{f-relxillnk}
\end{figure}

\begin{figure}[t]
\centering
\includegraphics[width=0.95\textwidth,trim=1.0cm 2.5cm 0.0cm 2.0cm,clip]{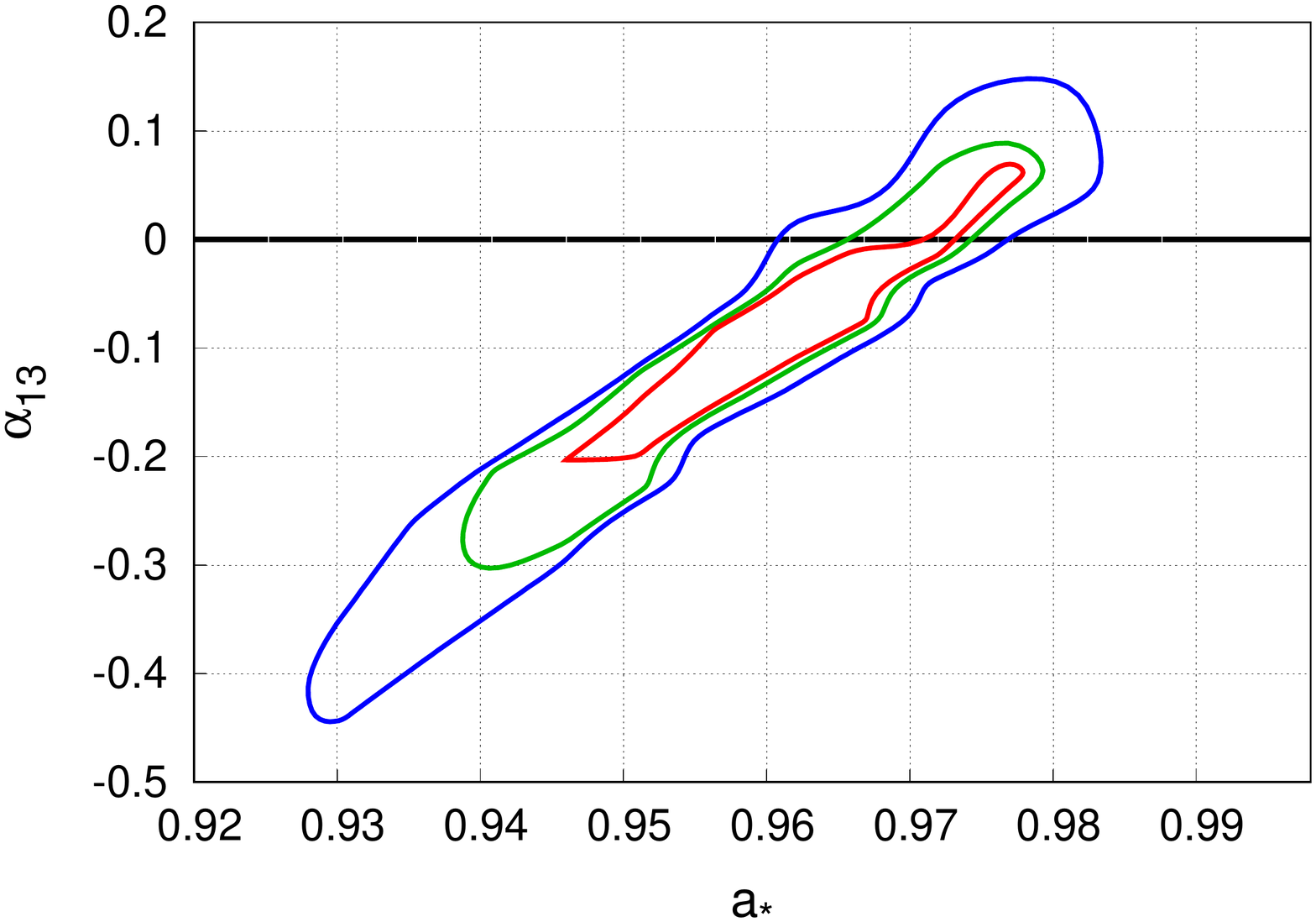}
\caption{Constraints on the spin parameter $a_*$ and the Johannsen deformation parameter $\alpha_{13}$ from the analysis of the reflection features in \textsl{XMM-Newton} and \textsl{NuSTAR} spectra of the supermassive black hole in MCG--06--30--15 reported in Ref.~\cite{Tripathi:2018lhx}. The red, green, and blue curves represent, respectively, the 68\%, 90\% and 99\% confidence level curves for two relevant parameters ($\Delta\chi^2 = 2.30$, 4.61, and 9.21, respectively). The horizontal black thick line at $\alpha_{13} = 0$ marks the Kerr solution. Figure from Ref.~\cite{Tripathi:2018lhx}.}
\label{f-mcg063015}
\end{figure}

Fig.~\ref{f-relxillnk} shows the impact of the Johannsen deformation parameter $\alpha_{13}$ on the reflection spectrum of thin disks assuming that the other model parameters do not change. The reflection spectrum has a more complicated structure than the thermal one and, in the presence of high-quality data, it is possible to break the parameter degeneracy. Fig.~\ref{f-mcg063015} shows the constraints on the black hole spin parameter $a_*$ and the Johannsen deformation parameter $\alpha_{13}$ from the analysis reported in Ref.~\cite{Tripathi:2018lhx} of \textsl{XMM-Newton} and \textsl{NuSTAR} data of the supermassive black hole in MCG--06--30--15. While we clearly see a correlation between the estimates of $a_*$ and $\alpha_{13}$, we have quite precise measurements of the two parameters, unlike the case shown in Fig.~\ref{f-lmcx1}.

\subsection{Other X-ray techniques}\label{subsec:other}

The continuum-fitting method and X-ray reflection spectroscopy are two mature techniques to test General Relativity with black hole X-ray data and have already provided robust constraints on new physics. However, there are even other X-ray techniques that have been discussed in the literature and may be used in the future. For the moment, these techniques cannot test General Relativity, either because we do not have yet a clear understanding of the underlying physical process or because we do not have yet the necessary observational data. This subsection lists these alternative X-ray techniques without entering the details, which can be found in the original papers.

\subsubsection{Quasi-periodic oscillations}

Quasi-periodic oscillations (QPOs) are narrow features at characteristic frequencies in the X-ray power density spectra of accreting black holes and neutron stars~\cite{Bambi:2017khi}. In the case of stellar-mass black holes in X-ray binaries, QPOs can be grouped into two classes: low-frequency QPOs (0.1-30~Hz) and high-frequency QPOs (40-450~Hz). There are three different types of low-frequency QPOs: type-A, type-B, and type-C. QPOs are observed even in the X-ray power density spectra of intermediate-mass black holes and supermassive black holes, but they are more difficult to detect due to insufficient observation lengths and modeling problems.

The exact origin of these QPOs is currently unknown, even if there are a number of proposals in the literature. In most models, the QPO frequencies are somehow related to the three fundamental frequencies of equatorial circular orbits of a test-particle: orbital frequency (the inverse of the orbital period), radial epicyclic frequency (the frequency of radial oscillations around the mean orbit), and vertical epicyclic frequency (the frequency of vertical oscillations around the mean orbit). Since these three frequencies depends on the spacetime metric, if we knew the correct relations between QPO frequencies and fundamental frequencies, the measurement of QPO frequencies could be used to measure the black hole spin parameter (if we assume General Relativity) or test the Kerr hypothesis.

Zdeněk Stuchl\'\i{}k was the first to propose to use QPOs to test the Kerr hypothesis in Ref.~\cite{Stuchlik:2008fy} and other authors have later extended the study of QPOs to other non-Kerr spacetimes~\cite{Johannsen:2010bi,Bambi:2012pa,Stefanov:2012fb,Aliev:2013jqz,Bambi:2013fea,Maselli:2014fca,Bambi:2016iip,Maselli:2017kic,Deligianni:2020tyz,Azreg-Ainou:2020bfl,Deligianni:2021ecz,Jiang:2021ajk}. The result strongly depends on the QPO model and therefore, without knowing which model is correct, we cannot get any robust measurement of the properties of a compact object with this technique. The method is promising for the future, because it is relatively easy to measure the QPO frequencies with good precision. However, depending on the QPO model and the spacetime metric, it may be difficult to simultaneously estimate the black hole spin and the deformation parameter because of a degeneracy between these two quantities.

\subsubsection{X-ray polarization}

The thermal and reflection spectra of thin accretion disks around black holes are expected to be partially polarized. The thermal spectrum would be initially unpolarized, but it is expected to become partially polarized because of Thomson scattering of X-ray photons off free electrons in the dense atmosphere of the accretion disk. The calculation of the polarization of the radiation emitted by an accretion disk requires the calculation of the photon trajectories, from the emission point in the disk to the detection point far from the source, and the parallel transport of the polarization vector along these photon trajectories. If we assume geodesic motion, this corresponds to calculate null geodesics of the background spacetime and the parallel transport of the polarization vector along these null geodesics. Preliminary studies to test the Kerr hypothesis from the analysis of the polarization of thermal spectra of accretion disks are reported in Refs.~\cite{Krawczynski:2012ac,Liu:2015ibq}. However, these tests require higher quality data than those available today.

\subsubsection{X-ray reverberation mapping}

\begin{figure}[t]
\centering
\includegraphics[width=0.95\textwidth,trim=4.0cm 4.0cm 0.0cm 1.0cm,clip]{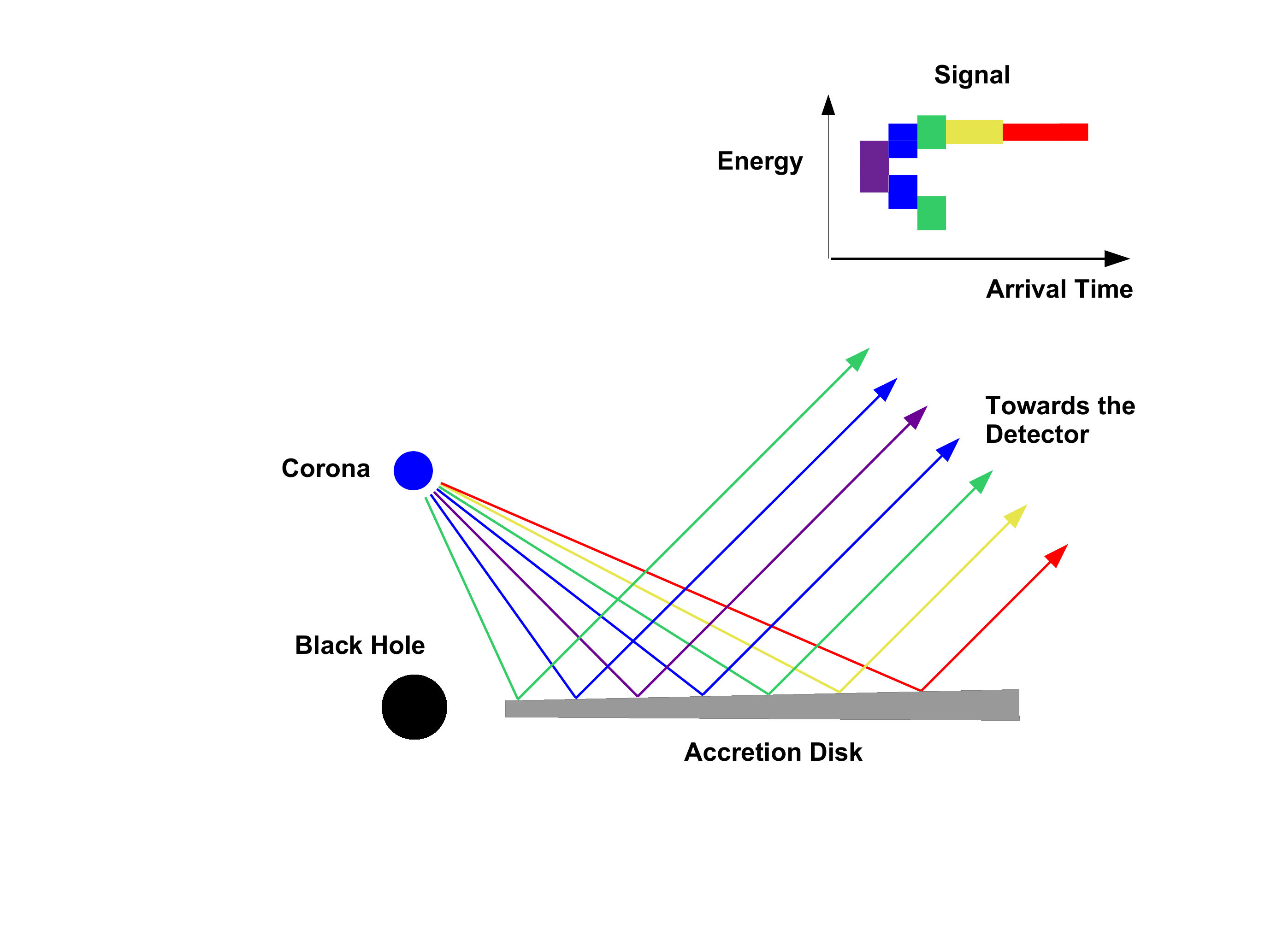}
\caption{When a new flare of the corona illuminates the disk, photons reflected from different points of the disk are detected at different times because the travel time depends on their exact path. Figure adapted from Ref.~\cite{Bambi:2017khi}.}
\label{f-reverberation}
\end{figure}

X-ray reflection spectroscopy is the analysis of the reflection spectrum of a disk integrated over a certain observational time. There is no information about the time arrival of the photons in the detectors. However, future X-ray observatories with high time resolutions and large effective areas should be able to measure the reflection spectrum as a function of time in response to a flare from the corona. The physical process is sketched in Fig.~\ref{f-reverberation}, where the corona is supposed to be a compact region along the spin axis of the black hole (lamppost corona): the corona has a new flare illuminating the disk, photons are reflected by the disk and detected by an instrument far from the source at different times depending on their reflection point on the disk. X-ray reverberation mapping refers to the analysis of the reflection spectrum as a function of time. Preliminary studies to test the Kerr hypothesis with X-ray reverberation mapping were reported in Refs.~\cite{Jiang:2014loa,Jiang:2016bdj}. Current X-ray observatories do not have a sufficiently large effective area, and thus any time resolved measurement is dominated by the intrinsic detection noise. With future X-ray observatories, X-ray reverberation mapping has the potentiality to constrain better the Kerr spacetime than X-ray reflection spectroscopy~\cite{Jiang:2014loa,Jiang:2016bdj}, but this requires a perfect knowledge of the coronal geometry, which is probably quite challenging.

\subsubsection{X-ray black hole imaging}

Current X-ray observatories have a limited angular resolution and every accreting black hole appears as a point-like X-ray source. However, future X-ray interferometry observatories will have the angular resolution to image thin accretion disks around black holes; see, for instance, Ref.~\cite{Uttley:2019ngm}. Today, we can image the supermassive black holes at the center of the Milky Way and at the center of the galaxy M87 at mm wavelengths with very long baseline interferometry facilities~\cite{EventHorizonTelescope:2019dse,EventHorizonTelescope:2022xnr}. X-ray interferometry missions will have the possibility of opening a new window to test General Relativity with black hole X-ray data, but it will probably take at least a few decades to reach the necessary technology for similar observations.


\section{Results}\label{sec:res}

The continuum-fitting method and X-ray reflection spectroscopy are two quite general techniques to probe the strong gravity region around astrophysical black holes and can potentially test a large number of scenarios beyond General Relativity. This section reviews the main results obtained so far, without entering technical details related to the analysis of specific sources, which can be found in the original publications.

Most of the studies have been devoted to test the Kerr hypothesis, following either a top-down or a bottom-up approach. In the latter case, so far the most widely used framework has been the Johannsen metric in Eq.~(\ref{eq-j}) and most studies reported in the literature have measured its deformation parameter $\alpha_{13}$ assuming that all other deformation parameters vanish. Any measurement consistent with $\alpha_{13} = 0$ confirms the Kerr hypothesis, while the detection of a possible non-vanishing $\alpha_{13}$ at a large confidence level value would indicate a violation of the Kerr hypothesis and General Relativity. As of now, there is no claim of any measurement of a non-vanishing $\alpha_{13}$ at a significant confidence level.

\begin{figure}[t]
\centering
\includegraphics[width=0.95\textwidth,trim=3.5cm 0.0cm 3.5cm 1.0cm,clip]{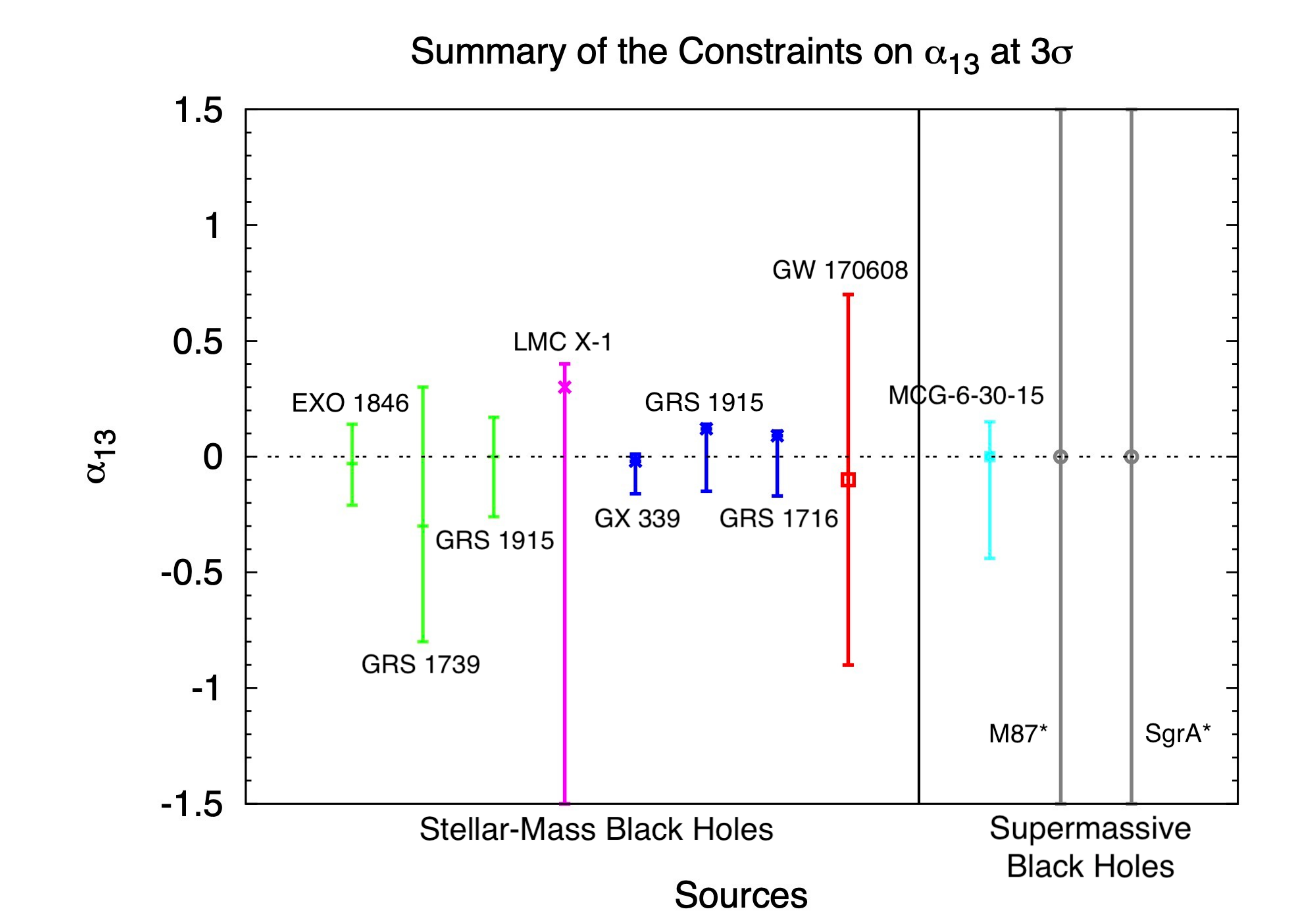}
\caption{Summary of the current constraints on the Johannsen deformation parameter $\alpha_{13}$ from X-ray reflection spectroscopy (green bars for stellar-mass black holes and cyan bar for supermassive black holes), continuum-fitting method (magenta bar), gravitational waves (red bar), and black hole imaging (gray bars). All constraints are at 3~$\sigma$. The horizontal dotted line at $\alpha_{13}=0$ marks the Kerr solution of General Relativity. The magenta and gray bars extend beyond the y-axis range. See the text for more details. Figure from Ref.~\cite{cb22}.}
\label{f-summary}
\end{figure}

Fig.~\ref{f-summary} compares the most stringent and robust constraints on $\alpha_{13}$ from different techniques. All constraints are at 3~$\sigma$ (statistical uncertainty only, but for all those constraints the systematic uncertainties are thought to be subdominant). Some constraints involving the analysis of the reflection features may be a bit different from those reported in the original publications because obtained from a more recent version of {\tt relxill\_nk}.
\begin{itemize}

\item The three green error bars are the three most stringent and robust constraints obtained from the analysis of the reflection features of stellar-mass black holes in X-ray binaries with {\tt relxill\_nk}. The constraints on $\alpha_{13}$ from EXO~1846 and GRS~1739 are obtained from the analyses of \textsl{NuSTAR} data in Ref.~\cite{Tripathi:2020yts}. The constraint from GRS~1915 is derived from the analysis of \textsl{Suzaku} data in Ref.~\cite{Zhang:2019ldz}. Other analyses of the reflection spectra of stellar-mass black holes have provided weaker and/or less robust constraints~\cite{Xu:2018lom,Zhang:2019zsn,Liu:2019vqh,Wang:2018bbr,Zhang:2020qbx,Liu:2021tyw}.

\item The magenta error bar is the constraint on $\alpha_{13}$ from the analysis of the thermal spectrum of the stellar-mass black hole in LMC~X-1 with {\tt nkbb} from \textsl{RXTE} data reported in Ref.~\cite{Tripathi:2020qco}. As we have already discussed in Subsection~\ref{subsec:cfm} and it is shown in Figs.~\ref{f-nkbb} and \ref{f-lmcx1}, there is a strong correlation between the measurements of the black hole spin parameter $a_*$ and the deformation parameter $\alpha_{13}$, and therefore the constraint on $\alpha_{13}$ is weak. For the same reason, we have not worked with other sources with a strong thermal component and no reflection features.

\item In some special cases, it is possible to test the Kerr hypothesis of the same source with both {\tt relxill\_nk} and {\tt nkbb}. This has been done only for three sources so far, but we can get the most stringent constraints on $\alpha_{13}$, as we can see from the blue error bars in Fig.~\ref{f-summary}. The constraint from GX~339 is derived from the analysis of a simultaneous observation of \textsl{NuSTAR} and \textsl{Swift} in which we see both a strong thermal component and very prominent reflection features with a very broadened iron line; see Ref.~\cite{Tripathi:2020dni} for the details. The constraint from GRS~1915 is obtained from the combined analysis of a \textsl{Suzaku} observation with strong reflection features and \textsl{RXTE} observations with a strong thermal component~\cite{Tripathi:2021rqs}. Last, the constraint from GRS~1716 is obtained from the analysis of three simultaneous observations \textsl{NuSTAR}+\textsl{Swift} in Ref.~\cite{Zhang:2021ymo}.

\item The red error bar is the most precise measurement of $\alpha_{13}$ from gravitational wave data in GWCT-1 and GWCT-2 obtained in Ref.~\cite{Shashank:2021giy} following the approach presented in Ref.~\cite{Cardenas-Avendano:2019zxd}. The measurement is obtained from the gravitational wave event GW170608 in which a $\sim$12~$M_\odot$ black hole merged to a $\sim$8~$M_\odot$ black hole to form a $\sim$18~$M_\odot$ black hole. See Ref.~\cite{Perkins:2022fhr,Riaz:2022rlx} for the caveats behind this constraints. The other events in GWCT-1 and GWCT-2 provide weaker constraints. In this specific case of the Johannsen deformation parameter $\alpha_{13}$, we see that the constraints from X-ray reflection spectroscopy are more stringent than those from gravitational waves, but this result should not lead to conclude that this is true for every deformation parameter. These two techniques are sensitive to different relativistic effects, and therefore it depends on the exact deformation from the Kerr solution which technique can provide the most stringent constraints. See, for instance, the case of other deformation parameters in Refs.~\cite{Abdikamalov:2021zwv,Shashank:2021giy,Yu:2021xen,Riaz:2022rlx}.

\item The cyan error bar is the most stringent and robust constraint on $\alpha_{13}$ from the analysis of the reflection spectrum of a supermassive black hole with {\tt relxill\_nk}. This measurement was obtained in Ref.~\cite{Tripathi:2018lhx} from the analysis of three simultaneous observations \textsl{NuSTAR}+\textsl{XMM-Newton} of the supermassive black hole in the galaxy MCG--06--30--15. While this constraint is comparable to the best constraints from the analysis of the reflection features of stellar-mass black holes, in general the latter are more suitable to test the Kerr hypothesis because brighter. However, MCG--06--30--15 is a very bright source, with a very prominent and broadened iron line, and the quality of those particular \textsl{NuSTAR} and \textsl{XMM-Newton} data is exceptionally good. Constraints on $\alpha_{13}$ from other analyses of reflection spectra of AGN have provided less stringent and/or less robust constraints; see Refs.~\cite{Tripathi:2018bbu,Choudhury:2018zmf,Tripathi:2019bya,Liu:2020fpv}

\item The two gray error bars refer to the constraints from the Event Horizon Telescope of, respectively, the supermassive black hole M87$^*$~\cite{EventHorizonTelescope:2020qrl}, at the center of the galaxy M87, and the supermassive black hole Sgr~A$^*$~\cite{EventHorizonTelescope:2022xqj}, at the center of the Milky Way. These constraints are currently significantly weaker than those from X-ray and gravitational wave data (their 3~$\sigma$ error bars extend beyond the y-axis range of Fig.~\ref{f-summary}). To reach the precision of the current constraints from X-ray and gravitational wave data, the Event Horizon Telescope would need to improve the current angular resolution roughly by an order of magnitude, which could be achieved by having one of the radio telescopes of the array in space. 

\end{itemize}

It is worth noting that, within an agnostic approach such as that of the Johannsen metric, we do not know if there is a correlation among the values of the deformation parameter of different sources. In general we can expect two main scenarios if the spacetime around astrophysical black holes is not described by the Kerr solution. In the first scenario, we can have a violation of the no-hair theorem: black holes are characterized by other parameters, and the values of these parameters can be different for different objects (just like the mass and the spin). In such a case, even if we tested the Kerr metric with high precision for a source, the result would not hold for other black holes. In the second scenario, black holes are not described by the Kerr solution, but some version of the no-hair theorem is still valid, so the value of the deformation parameter should be the same for all black holes. There may be even a third -- hybrid -- scenario, in which black hole have secondary hairs: black hokes are described by other parameters, but these new parameters are not really independent as they are determined by the mass and/or the spin. In such a case, we may be in the situation in which, for example, all stellar-mass black holes have a similar value of the deformation parameter, which, however, would be different from that of supermassive black holes.

{\tt relxill\_nk} has been used even to test specific gravity models in which black holes are not described by the Kerr solutions but motion of free particles is still described by the geodesic equations and the atomic physics in strong gravitational fields is the same as in our laboratories on Earth. The models studied so far include Einstein-Maxwell dilaton-axion gravity~\cite{Tripathi:2021rwb}, Kaluza-Klein gravity~\cite{Zhu:2020cfn}, asymptotically safe quantum gravity~\cite{Zhou:2020eth}, and conformal gravity~\cite{Zhou:2018bxk,Zhou:2019hqk}. The constraints obtained on these models are all consistent with the Kerr solution and are the most stringent among those reported in the literature so far.

In Ref.~\cite{Roy:2021pns}, we presented a modified version of {\tt relxill\_nk} to test the Weak Equivalence Principle, namely that any freely falling test-particle follows the geodesics of the spacetime, regardless of its internal structure and composition. Within the spirit of the bottom-up approach, we studied two cases. In the first model, photons followed null geodesics of the Kerr spacetime and massive particles in the disk followed time-like geodesics of the Johannsen spacetime with a possible non-vanishing deformation parameter $\alpha_{13}$. In the second model, massive particles in the disk followed time-like geodesics of the Kerr spacetime and photons followed null geodesics of the Johannsen spacetime with a possible non-vanishing deformation parameter $\alpha_{13}$. From the analysis of \textsl{NuSTAR} data of the stellar-mass black hole EXO~1846, we constrained the value of $\alpha_{13}$ for massive particles and the value of $\alpha_{13}$ for photons.

X-ray reflection spectroscopy can potentially test even the atomic physics in the strong gravitational fields of black holes. For instance, in Ref.~\cite{Davis:2016avf} the authors presented a model in which the effective value of the mass of elementary particles is determined by the value of a certain scalar field, which can be different near a black hole and on Earth. Atomic spectra near a black hole would be different from those in our laboratories on Earth because the value of the electron mass would be different. A very preliminary analysis on the possibility of measuring the value of the fine structure constant $\alpha$ in the strong gravitational field of a black hole from the study of the iron K$\alpha$ line was reported in Ref.~\cite{Bambi:2013mha}. A similar study of the whole reflection spectrum would be significantly more complicated, but it is certainly something worth investigating in the near future.


\section{Accuracy of gravity tests with black hole X-ray data}\label{sec:err}

After reviewing current observational constraints in the previous section, here we want to address a crucial question: are these constraints on the Kerr hypothesis from X-ray reflection spectroscopy robust? These measurements can certainly be precise, as shown in Fig.~\ref{f-summary}, but are they also accurate?

As discussed below, X-ray reflection spectroscopy can provide precise and accurate measurements of accreting black holes, but it is important to select the right sources and the right observations. Among the huge number of observations available in archive, only a very limited number of spectra are suitable to test the Kerr hypothesis. On the contrary, if we want to test the Kerr hypothesis for every source that presents a spectrum with blurred reflection features -- as it is often done in the case of black hole spin measurements under the assumption that General Relativity is correct -- it is very easy to get precise but non-accurate measurements.

\subsection{Selection of the spectra}

Theoretical reflection models have several parameters. If we want to measure all the parameters from the spectral fitting, we need high-quality data and a spectrum with strong signatures of relativistic effects. Without these two ingredients, we can unlikely break the parameter degeneracy of the model and we would be forced to freeze the values of some parameters in the fit, with the result that the final measurement may be biased.

To have high quality data, the source must be very bright to have a good statistics (assuming that our X-ray detector has no pile-up problems), the data should cover a wide energy band to see both the relativistically broadened iron line and the Compton hump, and we should have a good energy resolution at the iron line (which is the most informative part of the spectrum concerning relativistic effects). In this regard, \textsl{NuSTAR} observations of bright Galactic black holes are normally the most promising spectra to test the Kerr hypothesis~\cite{Tripathi:2020yts}, because the sources are bright, \textsl{NuSTAR} covers the energy band 3-78~keV, and its spectra are normally not affected by pile-up. In the case of simultaneous observations \textsl{NuSTAR}+\textsl{XMM-Newton} or \textsl{NuSTAR}+\textsl{NICER}, we can also benefit of a good energy resolution at the iron line~\cite{Tripathi:2018lhx}.

In order to have strong relativistic signatures in the reflection spectrum, it is necessary that the inner edge of the accretion disk is as close as possible to the black hole (which, in turn, requires to select very fast-rotating black holes with inner edge of the disk at the ISCO radius) and that the corona illuminates well the very inner part of the disk (which, in turn, requires that the corona is compact and very close to the black hole, so the strong light bending can focus most of the hard X-ray photons from the corona to the region around the inner edge of the disk). These two conditions lead to very broadened iron lines in reflection spectra~\cite{Dauser:2013xv}. In the presence of strong relativistic signatures in the reflection spectrum, it is possible to break the parameter degeneracy~\cite{Fabian:2014tda,Kammoun:2018ltv}.

Apart these two ``general'' requirements, it is also necessary to select sources with geometrically thin accretion disks. This condition can be satisfied if we select sources with an accretion luminosity between $\sim$5\% to $\sim$30\% of the Eddington limit~\cite{Steiner:2010kd}, but this require to have a reliable estimate of the black hole mass and distance. All the current reflection models employ thin accretion disks. The exact thickness of the disk is not relevant in the final measurement as long as the disks are thin, especially if we study very fast-rotating black holes as required to test the Kerr hypothesis~\cite{Abdikamalov:2020oci,Tripathi:2021wap}. On the contrary, if the black hole accretes from a thick disk, we can easily get precise but non-accurate measurements of the source~\cite{Riaz:2019bkv,Riaz:2019kat}.

For instance, the constraints on the Kerr metric reported in Ref.~\cite{Tripathi:2019bya} appear to be very precise, but they do not meet the conditions of high-quality data and geometrically thin disks, with the result that the systematics is not fully under control~\cite{Zhou:2019kwb}. The analysis is based only on data in the soft X-ray band, which do not cover the Compton hump. While there are large uncertainties on the masses and distances of those sources, some objects are thought to accrete near the Eddington limit, so the inner edge of the accretion disk is likely thick and any estimate of the black hole spin or of the deformation parameter can be affected by large systematic uncertainties~\cite{Riaz:2019bkv,Riaz:2019kat}.

If we assume that the inner edge is at the ISCO radius during the data analysis, it is also necessary to select sources that meet this condition. In the case of sources with a strong thermal component and an accretion luminosity between $\sim$5\% and $\sim$30\% of the Eddington limit, there is clear observational evidence that the inner edge is at the ISCO radius~\cite{Steiner:2010kd}. In other cases, this point should be explicitly checked during the data analysis. If the corona illuminates well the very inner part of the accretion disk, the fit can also measure the inner radius of the disk~\cite{Fabian:2014tda}. In general, this is not a problem for testing the Kerr hypothesis because we have to select in any case very fast-rotating black holes with the inner edge of the disk as close as possible to the compact object, so we obtain the same result if we impose that the inner edge is at the ISCO or if we leave the inner edge free in the fit~\cite{Tripathi:2020dni}.

\subsection{Accuracy of the theoretical models}

Apart the selection of the spectra, one may be worried about the accuracy itself of the theoretical models. Reflection models have been significantly developed in the past decade~\cite{Bambi:2020jpe}, but they still rely on a number of simplifications that may somewhat affect the final measurements. However, as it will be discussed below, with the available data from X-ray missions like \textsl{NuSTAR}, \textsl{NICER}, \textsl{Suzaku}, and \textsl{XMM-Newton}, uncertainties should be dominated by the statistical uncertainties rather than the systematic uncertainties once we have selected suitable spectra and chosen the correct theoretical models during the data analysis process.

The geometry of the corona is not yet well understood, while it is important in the calculation of reflection spectra because it determines the exact emissivity profile of the disk. However, the emissivity profiles produced by specific coronal geometries have been calculated in a number of studies~\cite{Wilkins:2011kt,Wilkins:2015nfa,Gonzalez:2017gzu,Riaz:2020svt}, and the conclusion is that a broken power law profile, or at most a twice broken power law profile as in Eq.~(\ref{eq-tbpl}), should be enough to approximate well the emissivity profile generated by a corona of arbitrary geometry. In the presence of high-quality data, the fit should be able to determine all parameters of this phenomenological emissivity profile. For example, in Ref.~\cite{Tripathi:2020yts} we have shown that the spectra that seem to require a lamppost corona can be also fit with a broken power law profile, always finding consistent estimates of the model parameters even if the fit is a bit worse (while the opposite is not true, namely the spectra that require a broken power law profile may not be fit with the less flexible lamppost model).

The accuracy of the disk model employed in reflection models has been recently tested with \textsl{NuSTAR} simulations of reflection spectra calculated from GRMHD-simulated disks in Ref.~\cite{Shashank:2022xyh}. The conclusion of that work is that we can recover the correct input parameters from the spectral analysis of the reflection features with current reflection models. The concept of inner edge of the disk is more likely replaced by the ``reflection edge'', which is still around the ISCO radius and separates the disk from the plunging region~\cite{Reynolds:2007rx}. The plunging region can be optically thick and produce reflection photons, but its density is so low that the gas is highly ionized: as a result, the reflection process is dominated by Compton scattering and the reflection spectrum looks like a power law without emission lines, so it does not affect the analysis of the reflection features and, in turn, the estimate of the model parameters~\cite{Reynolds:2007rx,Cardenas-Avendano:2020xtw}.

An effect completely ignored in current reflection analysis to test the Kerr hypothesis is the returning radiation, namely the radiation emitted by the disk and returning to the disk because of the strong light bending near a black hole~\cite{Riaz:2020zqb}. Such a radiation clearly alters the original emissivity profile produced by the corona~\cite{Dauser:2022zwc}. However, if we fit the emissivity with a broken power law or a twice broken power law, we should simply find the total emissivity profile rather than the original one produced by the illumination of the corona, with no impact on the measurements of the parameters. The reflection spectrum produced by the returning radiation is somewhat different from that produced by the direct radiation from the corona, since the returning radiation has a reflection spectrum and the direct radiation from the corona has a power law spectrum. However, the difference should be normally small because, apart the fluorescent emission lines and the Compton hump, even the reflection spectrum can be approximated well by a power law spectrum~\cite{Dauser:2022zwc}.


\section{Concluding remarks}\label{sec:c}

This chapter has reviewed the state-of-the-art of tests of General Relativity in the strong field regime with X-ray data of astrophysical black holes. As shown in Fig.~\ref{f-summary}, X-ray reflection spectroscopy can provide quite stringent tests of the Kerr hypothesis if compared with other methods and, as discussed in Section~\ref{sec:err}, these constraints are robust if we select suitable sources and observations. As of now, most efforts have been focused on testing the Kerr hypothesis. However, X-ray reflection spectroscopy can test even geodesic motion and the atomic physics in strong gravitational fields, which are not easily accessible to gravitational wave tests even in the future.

For the next few years, we cannot expect to significantly improve the current X-ray constraints in Fig.~\ref{f-summary}, as these are limited by the quality of the available X-ray data. The next generation of X-ray observatories, starting from the Sino-European mission \textsl{eXTP} currently scheduled to be launched in 2027~\cite{eXTP:2016rzs}, will provide unprecedented high-quality X-ray data. For that time, we have to further develop our current reflection models in order to reduce the systematic uncertainties related to the astrophysical set-up. For the future, we can also expect that other X-ray techniques will be able to provide interesting constraints on General Relativity, like the analysis of QPOs or X-ray reverberation mapping. For X-ray images of accreting black holes suitable to test General Relativity, we will probably have to wait for at least a few decades.


\begin{acknowledgement}
I thank Jerome Orosz for Fig.~\ref{f-bhxrb}, Jorge Casares and Jes\'us M. Corral-Santana for Fig.~\ref{f-bht}, and Matteo Guainazzi for Fig.~\ref{f-rmf}.
This work was supported by the National Natural Science Foundation of China (NSFC), Grant No.~12250610185 and Grant No. 11973019, the Natural Science Foundation of Shanghai, Grant No. 22ZR1403400, the Shanghai Municipal Education Commission, Grant No. 2019-01-07-00-07-E00035, and Fudan University, Grant No. JIH1512604.
\end{acknowledgement}






\end{document}